\documentclass[10pt,a4paper,twocolumn]{IEEEtran}

\usepackage{graphicx}
\usepackage[comma,numbers,square,sort&compress]{natbib}
\usepackage{epstopdf}
\usepackage{amsmath}
\usepackage{enumerate}
\usepackage{subfigure}
\usepackage{ntheorem}
\usepackage{amssymb}
\usepackage{mathrsfs}        % Include this line if your
\usepackage{indentfirst}

\newtheorem{definition}{Definition}
\newtheorem{lemma}{Lemma}
\newtheorem{theorem}{Theorem}
\newtheorem{corollary}{Corollary}
\newtheorem{remark}{Remark}

\begin{document}

\title{Global Structure Identifiability and Reconstructibility of an NDS with Descriptor Subsystems} % Title, preferably not more than 10 words.

\author{Tong Zhou and Kailin Yin% <-this % stops a space
\thanks{This work was supported in part by the NNSFC under Grant 61733008,  52061635102 and 61573209.}% <-this % stops a space
\thanks{Tong Zhou and Kailin Yin are with the Department of Automation, Tsinghua University, Beijing, 100084, P.~R.~China
        {(email: {\tt\small tzhou@mail.tsinghua.edu.cn, yinkl17@mails.tsinghua.edu.cn}.)}%
}}

\maketitle

\begin{abstract}                          % Abstract of not more than 200 words.    and causal/impulse free
This paper investigates requirements on a networked dynamic system (NDS) such that its subsystem interactions can be solely determined from experiment data or reconstructed from its overall model. The NDS is constituted from several subsystems whose dynamics are described through a descriptor form. Except regularity on each subsystem and the whole NDS, no other restrictions are put on either subsystem dynamics or subsystem interactions. A matrix rank based necessary and sufficient condition is derived for the global identifiability of subsystem interactions, which leads to several conclusions about NDS structure identifiability when there is some a priori information. This matrix also gives an explicit description for the set of subsystem interactions that can not be distinguished from experiment data only. In addition, under a well-posedness assumption, a necessary and sufficient condition is obtained for the reconstructibility of subsystem interactions from an NDS descriptor form model. This condition can be verified with each subsystem separately and is therefore attractive in the analysis and synthesis of a large-scale NDS. Simulation results show that rather than increases monotonically with the distance of subsystem interactions to the undifferentiable set, the magnitude of the external output differences between two NDSs with distinct subsystem interactions increases much more rapidly when one of them is close to be unstable. In addition, directions of probing signals are also very important in distinguishing external outputs of distinctive NDSs. These findings are expected to be helpful in identification experiment designs, etc.
\end{abstract}


\renewcommand{\labelenumi}{\rm\bf A\arabic{enumi})}

\section{Introduction}

Networked dynamic systems (NDS) arise in many areas of sciences and technologies, in which the overall behavior depends not only on the dynamics of each subsystem, but also largely on subsystem interactions \cite{Siljak1978,htw2009,pmssaxcas2010}. It is well known that the behaviors of the whole NDS may be significantly different from those of its subsystems due to their interactions. Subsystem interactions, known also as the structure of an NDS according to various literature, play an important role in NDS analysis and synthesis, including understanding generation mechanisms of a complex disease, distributed estimation and control, network security, etc. Under various situations, however, NDS structure information, including the existence and the strength of a direct effect between two subsystems, can hardly be obtained from NDS working principles and measurements. For instance, in a gene regulation network, measuring a direct influence among genes and other biomacromolecules is in general difficult and costly, and sometimes may even be impossible. As an another example, it is quite important to detect a structure change in a power system or a communication network, as an electricity transmission line or a communication channel may fail to work due to some accidents or attacks, that may cause significant economic losses and result in  great living inconveniences \cite{zg2012}. A frequently encountered example is that the structure of an autonomous vehicle system may vary with environments. And so on. All these situations make it necessary to estimate the structures of an NDS from experiment measurements. With simultaneous NDS scale increment and data aggregating in a social/financial/biological/manufacturing system, etc., developing an efficient NDS structure identification method attracts more and more research attentions from various different fields \cite{Koopmans1949,Manski1993,sd2020,zyl2018}.

Briefly, there are two major approaches in NDS dynamics descriptions. One treats each measured signal as a node and a transfer function from one measured signal to another measured signal as an edge \cite{hgb2019,sd2020,wvd2018}. The other treats each subsystem as a node and subsystem interactions as edges \cite{pmssaxcas2010,vtc2021,Zhou2020-1}. In either of these two approaches, the set of the edges connecting  different nodes of an NDS is called its structure, which can be associated to a graph that reveals direct influences among different signals or subsystems. While both approaches are wildly adopted in NDS analysis and synthesis, it is currently well known from the aspect of system structure information, that these two types of descriptions are different from either a transfer matrix function (TFM) model or a state space model \cite{sd2020}. More specifically, without additional information and/or other restrictions on the system, neither of their structures can be uniquely determined from a frequency/time domain model of the whole NDS. These imply that new theoretical issues arise in NDS structure identification which have been attracting extensive attentions in recent years, and various methods have been suggested. Some examples are \cite{pmssaxcas2010,sd2020,vtc2021,wvd2018,Zhou2020-1} and the references therein.

Specifically, \cite{sd2020} introduces a so-called dynamic structure function to deal with situations when there are some hidden system states and direct influences are to be estimated for part of the system states that can be directly measured. Some necessary and sufficient conditions have been established there for the ability of reconstructing these dynamical structure functions from the TFM of the whole NDS. In terms of eigenvectors, \cite{ptt2019} gives some necessary and sufficient conditions for detecting structure variations in an NDS with descriptor form subsystems and diffusive subsystem couplings. For an NDS with its subsystems being described by a state space model and coupled through their outputs, \cite{vtc2021} proven that its structure is identifiable only when the constant kernel of a TFM is exactly the zero vector, which is completely determined by the dynamics of each subsystem. It has also been proven there that under some special situations, this condition becomes also sufficient. When the dynamics of each subsystem is described by a state space model and its inputs/outputs are divided into internal and external ones, \cite{Zhou2020-1} proves that if in each subsystem, the TFM from its internal inputs to its external outputs is of full normal column rank (FNCR), while the TFM from its external inputs to its internal outputs is of full normal row rank (FNRR), then the structure of the NDS with arbitrary subsystem connections is identifiable. These restrictions have been partly removed in \cite{yz2021}, which gives a necessary and sufficient condition for NDS structure identifiability when only one of the aforementioned two assumptions is satisfied, as well as an explicit description for  the set of subsystem interactions that can not be differentiated from a particular subsystem connection matrix (SCM) through only experiment data.

In this paper, we investigate conditions on an NDS whose subsystem interactions can be identified from experiment data, under the situation that the dynamics of its subsystems are described by a descriptor form model. Both continuous and discrete time NDSs have been studied. Under the condition that the overall NDS and each of its subsystems are regular, it is shown that a specific SCM is globally identifiable, if and only if any other SCM leads to a different NDS TFM. This condition is further proven to be equivalent to that a constant matrix is of full column rank (FCR) that depends on this SCM and NDS subsystem dynamics. In addition, the right null space of this matrix gives a complete description for all the SCMs that can not be differentiated from this specific SCM using only experiment data. These results are extended to situations in which there are some a priori information about NDS subsystem interactions, and elements of the NDS SCM are not algebraically independent of each other. Subsystem interaction reconstructibility has also been investigated. It is proven that the SCM of an NDS can be completely recovered from a  descriptor form model of the whole NDS, when and only when two constant matrices that are associated with each subsystem independently, are respectively of FCR and of full row rank (FRR). This result is scalable for a large-scale NDS and appears helpful in  subsystem selections/designs. Some numerical simulation results have also been included, which show that rather than the distance of an SCM to the set of undifferentiable SCMs, it is the distance of an SCM to the set of the unstable NDS associated SCMs, that signifies the differences between the NDS external outputs associated with two distinctive SCMs. In addition, directions of probing signals are also very important in distinguishing NDS structures. These observations may inspire test signal selections and be helpful in identification experiment designs.

The remaining of this paper is organized as follows. Section 2 gives the NDS model adopted in this paper, a problem description, as well as  some preliminary results. Global structure identifiability of an NDS with some specific subsystem connections is investigated in Section 3, while some particular situations are studied in Section 4. Section 5 deals with subsystem interaction reconstruction from a descriptor form model of the whole NDS. A numerical example is given in Section 6 to illuminate the obtained theoretical results. Some concluding remarks are given in Section 7 with a brief  discussion about several further issues on NDS structure identification. Finally, an appendix is included that provides proofs of some technical results.

The following notation and symbols are adopted in this paper. $\mathcal{R}$ and $\mathcal{C}$ stand respectively for the set of real and complex numbers, $\mathcal{R}^m$, $\mathcal{R}^{m\times n}$ and $\mathcal{R}^{m\times n}[\lambda]$ respectively the set of $m$ dimensional real vectors, the set of $m\times n$ dimensional real matrices and the set of $m\times n$ dimensional matrix valued polynomials (MVP) with real coefficients and a finite degree. $[\star]_{ij}$ represents the $i$-th row $j$-th column element of a matrix, while $\star_{r}^\bot$/$\star_{l}^\bot$ the matrix whose columns/rows form a base of the right/left null space of a matrix. The subscript $r$ or $l$ is usually omitted when it is clear from the context. $diag\{X_i|^n_{i=1}\}$ stands for a diagonal matrix with its $i$-th diagonal block being $X_i$, while $col\{X_i|^n_{i=1}\}$ the vector/matrix stacked by $X_i|^n_{i=1}$ with its $i$-th row block vector/matrix being $X_i$, and $vec\{X\}$ the vector stacked by the columns of matrix $X$. $\rho_{max}(\star)$ and $\rho_{min}(\star)$ denote respectively the maximum and minimum absolute values of the eigenvalues of a square matrix, $\overline{\sigma}(\star)$ the maximum singular value of a matrix, and $det(\star)$ the determinant of a square matrix.
$deg x(\lambda)$ represents the degree of a polynomial $x(\lambda)$, $|\star|$ the number of elements in a set, and the superscript $T$ the transpose of a matrix/vector. $I_{m}$ and $0_{m\times n}$ stands for the $m$ dimensional identity matrix and the $m\times n$ dimensional zero matrix. The subscripts are often omitted when this piece of information is not very essential. The normal rank of a matrix valued function (MVF) is defined as its maximum rank when its variables vary over their  definition domains. When this rank is equal to the number of its rows/columns, this MVF is called full normal row/column rank (FNRR/FNCR).

\section{Problem Formulation and Preliminaries}\label{section2}

Consider the linear time invariant (LTI) NDS model adopted in \cite{Zhou2020-2}, in which subsystems are allowed to have distinctive dynamics and arbitrary direct interactions. Moreover, the dynamics of its subsystems are described by a descriptor form, which can approximately represent various actual systems in many fields like engineering, biology, economy, etc. \cite{Dai1989,Duan2010,Siljak1978}.

Specifically, for an NDS $\rm\bf\Sigma$ constituted from $N$ subsystems, the dynamics of its $i$-th subsystem $\Sigma_i$ is described as follows,
\begin{equation}\label{dyn}
\hspace*{-0.3cm}	\begin{bmatrix}\!
		E(i)\delta(x(t,i)) \\  z(t,i)\\  y(t,i) \!
	\end{bmatrix}
	\!=\!
	\begin{bmatrix}\!
		A_{xx}(i) & B_{xv}(i) & B_{xu}(i)\\
		C_{zx}(i) & D_{zv}(i) & D_{zu}(i)\\
		C_{yx}(i) & D_{yv}(i) & D_{yu}(i)  \!
	\end{bmatrix}\!\!
	\begin{bmatrix}\!
		x(t,i)\\v(t,i)\\u(t,i)\!
	\end{bmatrix}
\end{equation}
in which $E(i)$ is a real square matrix that may not be invertible. In actual applications, this matrix is usually utilized to reflect constraints on system variables, etc. $\delta(\star)$ denotes the derivative of a function with respect to time or a forward time shift operation, meaning that the above model can be either continuous time or discrete time. $t$ stands for the temporal variable, while $x(t,i)$ the state vector of the $i$-th subsystem $\Sigma_i$. Outputs and inputs of this subsystem are divided into internal and external parts, in which the internal ones are used to represent subsystem interactions, while the external ones are actual NDS inputs or outputs. In particular,  $u(t,i)$ and $y(t,i)$ are used to denote respectively the external input/output vectors of Subsystem $\Sigma_i$, while $v(t,i)$ and $z(t,i)$ respectively its internal input/output vectors, meaning signals obtained from other subsystems and signals sent to other subsystems.

In addition, subsystem interactions of the whole NDS $\rm\bf\Sigma$ are described by the following equation,
\begin{equation}\label{scm}
	v(t)=\Phi z(t)
\end{equation}
in which $v(t)$ and $z(t)$ are assembly expressions respectively for the internal input and output vectors of each subsystem. That is, $z(t)=col\{z(t,i)|^N_{i=1}\}$, $v(t)=col\{v(t,i)|^N_{i=1}\}$. Matrix $\Phi$ depicts interactions among NDS subsystems, which is called subsystem connection matrix (SCM). Considering each subsystem as a node and each nonzero element of $\Phi$ as a directed and weighted edge, a graph can be constructed, known as the topology or structure of the NDS $\rm\bf\Sigma$, representing direct interactions among NDS subsystems.

Throughout this paper, the dimension of a vector $\star(t,i)$ with $i=1,2,\cdots,N$ and $\star$ being $u$, $v$, $x$, $y$ or $z$, is denoted by   $m_{\star_i}$. Using these symbols, define an integer $m_\star$ as $m_\star=\sum\nolimits_{i=1}^Nm_{\star_i}$. Then the SCM $\Phi$ is clearly a $m_v\times m_z$ dimensional real matrix. Moreover, denote vectors $col\{x(t,i)|^N_{i=1}\}$, $col\{u(t,i)|^N_{i=1}\}$ and $col\{y(t,i)|^N_{i=1}\}$ respectively by $x(t)$, $u(t)$ and $y(t)$. To reveal that both the NDS $\rm\bf\Sigma$ and its external output vector $y(t)$ are dependent on its SCM $\Phi$, they are sometimes also written respectively as $\Sigma(\Phi)$ and $y(t,\Phi)$.

Concerning a lumped LTI plant, if its input-output relations can be described by the following equations,
\begin{equation}\label{D}
	E\delta(x(t))=Ax(t)+Bu(t),\quad y(t)=Cx(t)+Du(t)
\end{equation}
then this plant is called a descriptor system. Here $A, B, C, D$ and $E$ are constant real matrices with compatible dimensions. When the matrix $E$ is not invertible, this plant is sometimes also called a singular system. Compared with the extensively adopted state space model, this model is believed to be more natural and more convenient in expressing constraints among system variables and keeping structure information of plant dynamics \cite{Dai1989,Duan2010,Siljak1978,Zhou2020-2}.

Compared with a state space model, dynamic behaviors of a descriptor system are much more complicated. Special issues related to a descriptor system include that the associated equations may not have a solution, its response may not be unique to the same exciting signal under the same initial conditions, there may exist impulses in its response to a continuous exciting signal, its response may depend on future inputs, etc. Clearly, all these characteristics are not attractive to reveal the SCM $\Phi$ of the NDS $\rm\bf\Sigma$ from experiment data. To avoid occurrence of these phenomena in the adopted NDS model, some descriptor system related concepts are introduced.

\begin{definition}\label{defi0}
Assume that a descriptor system is described by Eq.(\ref{D}).
\begin{itemize}
\item Its initial state vector $x(0)$ and input vector $u(t)$ are called admissible, if there exists at least one trajectory $x(t)$ satisfying this equation.
\item This descriptor system is said to be regular, if there is a $\lambda\in\mathcal{C}$, such that $det(\lambda E-A) \neq 0$.
\end{itemize}
\end{definition}

Regularity and admissibility are special and important requirements for a descriptor system. A regular descriptor system has a unique state response and a unique output response, provided that it is stimulated by an admissible input under an admissible initial condition.

Throughout this paper, the following assumptions are adopted, which are called the regularity assumption for brevity in the remaining of this paper.
\begin{enumerate}
\item The whole NDS ${\rm\bf \Sigma}$, as well as each of its subsystems ${\rm\bf \Sigma}_{i}|_{i=1}^{N}$, is regular.
\end{enumerate}

The above arguments show that these assumptions are not very restrictive in actual applications, as both of them are necessary for the NDS ${\rm\bf \Sigma}$ to work properly.

To reveal an NDS structure from experiment data, a prerequisite is that it must be identifiable. Specifically, NDS structure identifiability can be defined as follows.

\begin{definition}\label{defi1}
Concerning the NDS $\rm\bf\Sigma$ of Eq.(\ref{dyn})-(\ref{scm}), assume that its SCM $\Phi$ belongs to a prescribed set $\mathscr{S}$.
\begin{itemize}
\item Two distinctive SCMs $\Phi \in \mathscr{S}$ and $\widetilde{\Phi} \in \mathscr{S}$ are said to be differentiable, if for an arbitrary admissible initial state vector $x(0)$, there exists at least one external admissible input signal $u(t)|_{t=0}^{\infty}$, such that the external output $y(t,\Phi)|_{t=0}^{\infty}$ of the NDS $\Sigma(\Phi)$ is different from the external output $y(t,\widetilde{\Phi})|_{t=0}^{\infty}$ of the NDS $\Sigma(\widetilde{\Phi})$. Otherwise, these two SCMs are called undifferentiable.
\item The structure of the NDS $\rm\bf\Sigma$ is called globally identifiable at a specific SCM $\Phi_0$, if it is differentiable  from any other SCM $\Phi \in \mathscr{S}$. Otherwise, this SCM $\Phi_0$ is called globally unidentifiable.
\item The structure of the NDS $\rm\bf\Sigma$ is called locally identifiable at a specific SCM $\Phi_0$, if there exists an $\varepsilon$-neighborhood $\mathscr{B}(\Phi_0,\varepsilon)$ which is a subset of $\mathscr{S}$, such that it is differentiable from any other SCM $\Phi\in\mathscr{B}(\Phi_0,\varepsilon)$. Otherwise, this SCM $\Phi_0$ is called locally unidentifiable.
\item The structure of the NDS $\rm\bf\Sigma$ is called locally/globally identifiable, if it is locally/globally identifiable at almost every $\Phi_0 \in \mathscr{S}$.
\end{itemize}
\end{definition}

Clearly, NDS structure identifiability is a particular situation of parameter identifiability. The latter has been attracting extensive research attentions for a long time in various fields, including engineering, finance, biology, etc. Nevertheless, it is still a challenging issue even for a lumped LTI system \cite{adm2020,Koopmans1949,Ljung1999,wvd2018,Zhou2020-1}. As structure information is widely recognized to be quite important in NDS analysis and synthesis, structure identifiability is used throughout this paper, in order to reflect this importance and to distinguish it from the traditional parameter identifiability, which is also adopted in other works, such as \cite{ptt2019,sd2020,wvd2018,vtc2021} and the references therein.

From this definition, it is clear that when the structure of the NDS $\rm\bf\Sigma$ is not identifiable at a particular SCM $\Phi_0$, whether it is locally or globally, then no matter what external exciting signals is used, how long an experiment is performed, and how advanced that an identification algorithm is, this set of subsystem interactions cannot be uniquely determined. This means that structure identifiability is an intrinsic property held by an NDS and has no concerns with data.

A closely related concept is structure reconstructibility \cite{sd2020}, which is defined as follows.

\begin{definition}\label{defi3}
An estimated model of the NDS ${\rm\bf\Sigma}$ is said to be consistent with its structure, if there exists an SCM $\Phi$ such that this model can have a realization of Eqs.(\ref{dyn}) and (\ref{scm}). In addition, if for each consistent model, there is only one SCM $\Phi$ that leads to this model, then the NDS structure is called reconstructible.
\end{definition}

The major objectives of this paper are to find computationally attractive conditions such that the SCM $\Phi$ can be identified from external input-output data of the NDS $\rm\bf\Sigma$, as well as those for the reconstructibility of the SCM $\Phi$ from a time domain model of the whole NDS $\rm\bf\Sigma$.

It is worthwhile to mention that if there are some subsystem parameters that are to be estimated, then through introducing some auxiliary internal subsystem inputs and outputs, it is possible to include them into the SCM $\Phi$. Details can be found in \cite{Zhou2020-2}. This means that the results of this paper are also applicable to parameter estimations for an NDS subsystem.

\section{NDS Structure Identifiability}\label{section3}

This section investigates structure identifiability verifications for the NDS $\rm\bf\Sigma$ of Section $\ref{section2}$. To make mathematical derivations more concise, the following matrices are defined. $E=diag\{E(i)|^N_{i=1}\}$, $A_{xx}=diag\{A_{xx}(i)|^N_{i=1}\}$, $B_{\star\#}=diag\{B_{\star\#}(i)|^N_{i=1}\}$, $C_{\star\#}=diag\{C_{\star\#}(i)|^N_{i=1}\}$, $D_{\star\#}=diag\{D_{\star\#}(i)|^N_{i=1}\}$, in which $\star=x,y$ or $z$, $\#=x,u$ or $v$.

Substitute Eq.(\ref{scm}) into Eq.(\ref{dyn}), the dynamics of the NDS $\rm\bf\Sigma$ can be equivalently rewritten as follows using the aforementioned symbols,
\begin{eqnarray}
& & \hspace*{-0.20cm}\begin{bmatrix}
		E & 0 \\ 0 & 0
\end{bmatrix}\!\!\!
\begin{bmatrix}
 \delta(x(t)) \\ \delta(z(t))
\end{bmatrix}
\!\!=\!\!	\begin{bmatrix}
		\! A_{xx} & B_{xv}\Phi \\
		C_{zx} & D_{zv}\Phi \!-\! I_{m_z}
	\!\end{bmatrix}\!\!\!
    \begin{bmatrix}
		x(t) \\ z(t)
	\end{bmatrix}
\!\!+\!\!	\begin{bmatrix}
		B_{xu} \\ D_{zu}
	\end{bmatrix}\!\! u(t)
\label{SSM-x-1} \\
& &\hspace*{-0.20cm} y(t)=\begin{bmatrix}
		C_{yx} & D_{yv}\Phi
	\end{bmatrix}
	\begin{bmatrix}
		x(t) \\ z(t)
	\end{bmatrix}
+  D_{yu} u(t)
\label{SSM-x-2}
\end{eqnarray}
in which the zero matrices have a compatible but usually different dimensions. Obviously, these expressions take completely the same form of a descriptor system given by Eq.(\ref{D}).

For an NDS subsystem ${\rm\bf\Sigma}_{i}$ with $i=1,2,\cdots,N$, define TFMs $G_{zu}(\lambda,i)$, $G_{zv}(\lambda,i)$, $G_{yu}(\lambda,i)$ and $G_{yv}(\lambda,i)$ respectively as
\begin{multline}\label{tfms}
	\begin{bmatrix}
		G_{yu}(\lambda,i) & G_{yv}(\lambda,i)\\
		G_{zu}(\lambda,i) & G_{zv}(\lambda,i)
	\end{bmatrix}
	=
	\begin{bmatrix}
		D_{yu}(i) & D_{yv}(i)\\
		D_{zu}(i) & D_{zv}(i)
	\end{bmatrix}
	+
	\begin{bmatrix}
		C_{yx}(i)\\C_{zx}(i)
	\end{bmatrix}\\
	\times
	[\lambda E(i)-A_{xx}(i)]^{-1}
	\begin{bmatrix}
		B_{xu}(i) & B_{xv}(i)
	\end{bmatrix}
\end{multline}
in which $\lambda$ stands for the Laplace transformation variable $s$ for a continuous time NDS or the $\mathcal{Z}$ transformation variable $z$ for a  discrete time NDS. Recall that each NDS subsystem is assumed to be regular, which means that the matrix pencil $\lambda E(i)-A_{xx}(i)$ is invertible for each $i=1,2,\cdots,N$. Hence, the above TFMs are well defined. Using these TFMs, define further a block diagonal TFM   $G_{\star\#}(\lambda)$ as $G_{\star\#}(\lambda)=diag\{G_{\star \#}(\lambda,i)\vert^N_{i=1}\}$, in which $\star=z$ or $y$ and $\#=u$ or $v$.

On the other hand, note that when the regularity assumption is satisfied, the matrix pencil $\lambda E -A_{xx}$ is also invertible due to its block diagonal structure. Based on this observation, direct matrix manipulations show that,
\begin{eqnarray}
& & det\left\{\lambda
\begin{bmatrix}
		E & 0 \\ 0 & 0
\end{bmatrix} -\begin{bmatrix}
		\! A_{xx} & B_{xv}\Phi \\
		C_{zx} & D_{zv}\Phi \!-\! I_{m_z}
	\!\end{bmatrix} \right\} \nonumber\\
&=& det\left(\lambda E - A_{xx}\right)\times det\left\{I_{m_z} - G_{zv}(\lambda)\Phi\right\} \nonumber\\
&=& det\left(\lambda E - A_{xx}\right)\times det\left\{I_{m_v} - \Phi G_{zv}(\lambda)\right\}
\label{Regularity}
\end{eqnarray}
It can therefore be declared that under the regularity assumption, the inverses of both the TFMs $I_{m_v}-\Phi G_{zv}(\lambda)$ and $I_{m_z}- G_{zv}(\lambda) \Phi$ are also well defined. This means that the following TFM $H(\lambda,\Phi)$ of the whole NDS $\Sigma(\Phi)$ is also well defined,
\begin{equation}\label{tfm}
	H(\lambda,\Phi)=G_{yu}(\lambda)+G_{yv}(\lambda)[I_{m_v}-\Phi G_{zv}(\lambda)]^{-1}\Phi G_{zu}(\lambda)
\end{equation}

On the basis of these TFMs, the following results are obtained which take completely the same form as those of Theorem 1 in \cite{Zhou2020-1}.

\begin{lemma}\label{lemma2}
Assume that the NDS $\rm\bf\Sigma$ satisfies Assumption A1). Then its structure is globally identifiable, if and only if for every two different SCMs $\Phi$ and $\Phi_0$ in the set $\mathscr{S}$, the corresponding NDS TFMs satisfy  $H(\lambda,\Phi)\not\equiv  H(\lambda,\Phi_0)$.
\end{lemma}

\noindent\textbf{Proof:} Perform Laplace/${\mathcal Z}$ transformation on both sides of Eqs.(\ref{dyn}) and (\ref{scm}) when $\delta(\cdot)$ is the derivative of a function with respect to time/a forward time shift operation. Note that when the regularity assumption is satisfied, both the TFM $I_{m_v}-\Phi G_{zv}(\lambda)$ and the matrix pencil $\lambda E(i) - A_{xx}(i)$ with $i\in\{1,2,\cdots,N\}$ are invertible. Then  arguments similar to the proof of Theorem 1 in \cite{Zhou2020-1} lead to the conclusions. This completes the proof.   \hspace{\fill}$\Diamond$

From this lemma, it is also clear that if there exists a particular SCM $\Phi_0\in \mathscr{S}$, such that for any SCM $\Phi\neq\Phi_0$ belonging to the set $\mathscr{S}$, the associated NDS TFM $H(\lambda,\Phi)$ is different from the TFM $H(\lambda,\Phi_0)$, then the NDS $\rm\bf\Sigma$ is globally identifiable at this specific SCM $\Phi_0$. Similar conclusions can be stated for local structure identifiability. These observations also reveal that for the adopted NDS model, its structure identifiability is equivalent to its reconstructibility from a TFM model.

While the above lemma gives a frequency domain criterion for NDS  structure identifiability, it is still not computationally verifiable. To overcome this difficulty, some knowledge on TFM decompositions, such as the Smith-McMillan form, coprime factorization, etc., are required, which are well known and have been playing important roles in system and control theories \cite{Kailath1980,zdg1996}.

In the remaining of this section, it is assumed that there is not any a priori information about NDS subsystem interactions. That is, each element of the SCM $\Phi$ can take an arbitrary real value. Several well encountered situations are discussed in the following Section 4, in which some a priori information is available for this SCM.

At first, we investigate NDS structure identifiability under the following assumptions.
\begin{enumerate}
\setcounter{enumi}{1}
\item There exists at least one integer pair $(i,\;j)$ with $1\leq i,j\leq N$, such that the TFM $G_{zu}(\lambda,i)$ is not of FNRR, while the TFM $G_{yv}(\lambda,j)$ is not of FNCR.
\end{enumerate}

Under this hypothesis, denote by $r^{[i]}_{yv}$ the normal rank of the TFM $G_{yv}(\lambda,i)$ for each $i=1,\cdots,N$. Note that each TFM $G_{yv}(\lambda,i)$ is a rational MVF. It has a  Smith-McMillan form that can be written as follows,
\begin{equation}\label{SM}
G_{yv}(\lambda,i) \!=\! U_{yv}(\lambda,i)
	\!\! \left[\!\begin{array}{cc}
diag\{\kappa^{[j]}_{yv}(\lambda,i)|^{r^{[i]}_{yv}}_{j=1}
\} & 0 \\ 0 & 0
	\end{array} \!\right] \!\!
	V^{T}_{yv}(\lambda,i)
\end{equation}
in which the zero matrices have compatible but usually different dimensions, while $U_{yv}(\lambda,i) \in \mathcal{R}^{m_{y_i}\times m_{y_i}}[\lambda]$ and $V_{yv}(\lambda,i) \in \mathcal{R}^{m_{v_i}\times m_{v_i}}[\lambda]$ are unimodular. Moreover, $\kappa^{[j]}_{yv}(\lambda,i)|^{r^{[i]}_{yv}}_{j=1}$ are real rational functions that are not identically equal to zero and have a finite degree.

Note that for each $i=1,2,\cdots,N$, the $m_{z_i}\times m_{v_i}$ dimensional TFM $G_{zv}(\lambda,i)$ is also a rational MVF. There certainly exist a $N_{zv}(\lambda,i)\in\mathcal{R}^{m_{z_i}\times m_{v_i}}[\lambda]$ and a $D_{zv}(\lambda,i)\in\mathcal{R}^{m_{v_i}\times m_{v_i}}[\lambda]$ which are right coprime and satisfy $G_{zv}(\lambda,i)=N_{zv}(\lambda,i)D^{-1}_{zv}(\lambda,i)$. An expression like this is widely known as a right matrix fraction description (MFD) of a TFM \cite{Kailath1980,zdg1996}.

For each unimodular MVP $V_{yv}(\lambda,i)$, denote its inverse by $V_{yv}^{[iv]}(\lambda,i)$.  As $V_{yv}(\lambda,i)$ is an unimodular, we have that its inverse $V_{yv}^{[iv]}(\lambda,i)$, and therefore $V_{yv}^{[iv]T}(\lambda,i)$, is also a MVP. This means that $D^{-1}_{zv}(\lambda,i)V^{[iv]T}_{yv}(\lambda,i)$ is a $m_{v_i}\times m_{v_i}$ dimensional rational fractional matrix (RFM). Hence, there exist $R(\lambda,i), Q(\lambda,i)$ and $\Omega(\lambda,i)\in\mathcal{R}^{m_{v_i}\times m_{v_i}}[\lambda]$, such that the MVPs $Q(\lambda,i)$ and $\Omega(\lambda,i)$ are right coprime, and
\begin{displaymath}
D^{-1}_{zv}(\lambda,i)V_{yv}^{[iv]T}(\lambda,i) = R(\lambda,i) + Q(\lambda,i)\Omega^{-1}(\lambda,i)
\end{displaymath}
Moreover, $Q(\lambda,i)\Omega^{-1}(\lambda,i)$ is a strict RFM.

With these symbols, define MVPs $X(\lambda)$ and $Y(\lambda)$ respectively as
\begin{align}
\hspace*{-0.25cm}X(\lambda)&\!=\!diag\{D_{zv}(\lambda,i)[R(\lambda,i)\Omega(\lambda,i) \!+\! Q(\lambda,i)]|^N_{i=1}\}
	\\
\hspace*{-0.25cm}Y(\lambda)&\!=\! diag\{N_{zv}(\lambda,i)[R(\lambda,i)\Omega(\lambda,i) \!+\! Q(\lambda,i)]|^N_{i=1}\}
\end{align}
Obviously, both of these two MVPs are block diagonal and have a finite degree. On the basis of these two MVPs, the following results are  derived, which give a necessary and sufficient condition for the global structure identifiability of the NDS $\rm\bf\Sigma$ at a specific SCM, say $\Phi_0$. Their proof is deferred to the appendix.

\begin{theorem}\label{theo1}
Assume that the NDS $\rm\bf\Sigma$ simultaneously satisfies the regularity assumption A1) and Assumption A2). Then this NDS is globally identifiable at a specific SCM $\Phi_0$, if and only if for any nonzero real vector $\delta\in\mathcal{R}^{m_v}$, the following equation does not have a solution,
	\begin{equation}\label{lemma}
		[X(\lambda)-\Phi_0Y(\lambda)]\alpha(\lambda)=\delta
	\end{equation}
in which $\alpha(\lambda)$ is a vector valued polynomial (VVP) that may have an infinite degree.
\end{theorem}

It is worthwhile to mention that for each subsystem ${\rm\bf\Sigma}_i$ with $1\leq i\leq N$, the right MFDs $N_{zv}(\lambda,i)D^{-1}_{zv}(\lambda,i)$ and $Q(\lambda,i)\Omega^{-1}(\lambda,i)$ are not unique. But this nonuniqueness does not affect the conclusions of Theorem $\ref{theo1}$.

To be specific, for each $i=1,2,\cdots,N$, let $W_1(\lambda,i)\in\mathcal{R}^{m_{v_i}\times m_{v_i}}[\lambda]$ be an arbitrary unimodular MVP. Define two MVPs $\widetilde{N}_{zv}(\lambda,i)$ and $\widetilde{D}_{zv}(\lambda,i)$ respectively as $\widetilde{N}_{zv}(\lambda,i)=N_{zv}(\lambda,i)W_1(\lambda,i)$ and  $\widetilde{D}_{zv}(\lambda,i)=D_{zv}(\lambda,i)W_1(\lambda,i)$. Then $\widetilde{N}_{zv}(\lambda,i)\widetilde{D}^{-1}_{zv}(\lambda,i)$ is also a right MFD of the TFM $G_{zv}(\lambda,i)$. Accordingly, there exist
$\widetilde{R}(\lambda,i)$, $\widetilde{Q}(\lambda,i)$ and $\widetilde{\Omega}(\lambda,i)$ belonging to $\mathcal{R}^{m_{v_i}\times m_{v_i}}[\lambda]$, such that
$\widetilde{Q}(\lambda,i)$ and $\widetilde{\Omega}(\lambda,i)$ are right coprime, and $\widetilde{D}^{-1}_{zv}(\lambda,i)V_{yv}^{[iv]T}(\lambda,i)=\widetilde{R}(\lambda,i)+\widetilde{Q}(\lambda,i)\widetilde{\Omega}^{-1}(\lambda,i)$. Define MVPs $\widetilde{X}(\lambda)$ and $\widetilde{Y}(\lambda)$ respectively as
\begin{align*}
	\widetilde{X}(\lambda)&= diag\{\widetilde{D}_{zv}(\lambda,i)[\widetilde{R}(\lambda,i)\widetilde{\Omega}(\lambda,i)+
	\widetilde{Q}(\lambda,i)]|^N_{i=1}\}\\
	\widetilde{Y}(\lambda)&=diag\{\widetilde{N}_{zv}(\lambda,i)[\widetilde{R}(\lambda,i)\widetilde{\Omega}(\lambda,i)+
	\widetilde{Q}(\lambda,i)]|^N_{i=1}\}
\end{align*}
On the other hand, $\widetilde{D}^{-1}_{zv}(\lambda,i)V_{yv}^{[iv]T}(\lambda,i) = W_1^{-1}(\lambda,i)\times$ $\left[{R}(\lambda,i) + {Q}(\lambda,i){\Omega}^{-1}(\lambda,i) \right]$. It can be straightforwardly shown that
\begin{eqnarray*}
& & \widetilde{R}(\lambda,i)\widetilde{\Omega}(\lambda,i)+
	\widetilde{Q}(\lambda,i)  \\
&=& W_1^{-1}(\lambda,i)\left[{R}(\lambda,i){\Omega}(\lambda,i)+ {Q}(\lambda,i)\right] {\Omega}^{-1}(\lambda,i)\widetilde{\Omega}(\lambda,i)
\end{eqnarray*}
We therefore have that $\widetilde{X}(\lambda)\equiv X(\lambda)\Xi(\lambda)$ and $\widetilde{Y}(\lambda)\equiv Y(\lambda)\Xi(\lambda)$, in which $\Xi(\lambda) = diag\{ {\Omega}^{-1}(\lambda,i)\widetilde{\Omega}(\lambda,i)|^N_{i=1}\}$. Apparently, the MVF $\Xi(\lambda)$ is block diagonal and invertible, and therefore does not influence the existence of a solution to Eq.(\ref{lemma}). That is, the nonuniqueness of the right MFD for the TFM $G_{zv}(\lambda,i)$ with $i=1,2,\cdots,N$, does not affect the conclusions of Theorem $\ref{theo1}$.

On the other hand, for each $i=1,2,\cdots,N$, let $W_2(\lambda,i)$ be another arbitrary $m_{v_i}\times m_{v_i}$ dimensional unimodular MVP. Define MVPs $\overline{Q}(\lambda,i)$ and $\overline{\Omega}(\lambda,i)$ respectively as $\overline{Q}(\lambda,i)=Q(\lambda,i)W_2(\lambda,i)$ and   $\overline{\Omega}(\lambda,i)=\Omega(\lambda,i)W_2(\lambda,i)$. Then $\overline{Q}(\lambda,i)\overline{\Omega}^{-1}(\lambda,i) = Q(\lambda,i)\Omega^{-1}(\lambda,i)$, and $\overline{Q}(\lambda,i)$ and $\overline{\Omega}^{-1}(\lambda,i)$ are right coprime. Define MVPs $\overline{X}(\lambda)$ and $\overline{Y}(\lambda)$ respectively as
\begin{align*} \overline{X}(\lambda)&=diag\{D_{zv}(\lambda,i)[R(\lambda,i)\overline{\Omega}(\lambda,i)+
	\overline{Q}(\lambda,i)]|^N_{i=1}\}\\ \overline{Y}(\lambda)&=diag\{N_{zv}(\lambda,i)[R(\lambda,i)\overline{\Omega}(\lambda,i)+
	\overline{Q}(\lambda,i)]|^N_{i=1}\}
\end{align*}
We further have that $[\overline{X}(\lambda)-\Phi_0 \overline{Y}(\lambda)]\alpha(\lambda)\equiv[X(\lambda)-\Phi_0Y(\lambda)]\overline{\alpha}(\lambda)$, in which $\overline{\alpha}(\lambda)=diag\{W_2(\lambda,i)|^N_{i=1}\}\alpha(\lambda)$. Hence, if ${\alpha}(\lambda)$ is a VVP that may have an infinite degree, then $\widehat{\alpha}(\lambda)$ also has this property, and vice versa, noting that for each $i=1,2,\cdots,N$, $W_2(\lambda,i)$ is unimodular. This means that the nonuniqueness of the right MFD  $Q(\lambda,i)\Omega^{-1}(\lambda,i)$ does not affect the conclusions of Theorem $\ref{theo1}$ either.

To get a criterion for NDS structure identifiability that can be computationally verified, denote the normal rank of the MVP $X(\lambda)-\Phi_0 Y(\lambda)$ by $r(\Phi_0)$. Write the  Smith form of this MVP as
\begin{equation}\label{Smith}
	X(\lambda)-\Phi_0Y(\lambda)=U_{\Phi_0}(\lambda)
	\left[\begin{array}{cc}
diag \{\mu^{[j]}_{\Phi_0}(\lambda)|^{r(\Phi_0)}_{j=1}\} & 0\\0 & 0
	\end{array} \right]
	V^T_{\Phi_0}(\lambda)
\end{equation}
in which the zero matrices have compatible but generally different dimensions, while $U_{\Phi_0}(\lambda), V_{\Phi_0}(\lambda)\in \mathcal{R}^{m_{v}\times m_{v}}[\lambda]$ are unimodular. Moreover, $\mu^{[j]}_{\Phi_0}(\lambda)\vert^{r(\Phi_0)}_{j=1}$ are real coefficient polynomials that are not identically equal to zero and have a finite degree. Denote the inverse of the unimodular MVP $U_{\Phi_0}(\lambda)$ by  $U^{[iv]}_{\Phi_0}(\lambda)$, and partition it into two row blocks as follows,
\begin{equation}
	U^{[iv]}_{\Phi_0}(\lambda)=
	\begin{bmatrix}
		U^{[iv,1]}_{\Phi_0}(\lambda) \\ U^{[iv,2]}_{\Phi_0}(\lambda)
	\end{bmatrix}
\end{equation}
in which the sub-MVP $U^{[iv,1]}_{\Phi_0}(\lambda)$ has $r(\Phi_0)$ rows. Let $p(\Phi_0)$ denote the degree of the MVP $U^{[iv,2]}_{\Phi_0}(\lambda)$. That is, $p(\Phi_0) = \max_{i,j}\{deg[U^{[iv,2]}_{\Phi_0}(\lambda)]_{ij}\}$. Then there exist some real matrices $U^{[iv,2]}_{\Phi_0,k}$ with $k=0,1,\cdots,p(\Phi_0)$,  such that $U^{[iv,2]}_{\Phi_0}(\lambda)=\sum^{p(\Phi_0)}_{k=0} \lambda^k U^{[iv,2]}_{\Phi_0,k}$. On the basis of this expression of the MVP $U^{[iv,2]}_{\Phi_0}(\lambda)$, the following results are obtained for NDS structure identifiability. Their proof is postponed to the appendix.

\begin{theorem}\label{theo2}
Assume that the NDS $\rm\bf\Sigma$ satisfies both Assumptions A1) and A2). Then its structure is globally identifiable at a specific SCM $\Phi_0$, if and only if the matrix $U^{[iv,2]}_{\Phi_0} = col\{ {U^{[iv,2]}_{\Phi_0,k}} |^{p(\Phi_0)}_{k=0}\}$ is of FCR.
\end{theorem}

\begin{remark}
This theorem clarifies that global structure identifiability of the NDS $\rm\bf\Sigma$ at some specific subsystem connections can be transformed into an algebraic criterion whose verification is in principle computationally possible, especially when the NDS does not have a great number of subsystems. Denote the dimension of the right null space of the matrix $U^{[iv,2]}_{\Phi_0}$ by $\widehat{r}(\Phi_0)$. Moreover, define a set $\mathscr{U}(\Phi_0)$ as follows,
	\begin{equation}\label{unidarea}
		\mathscr{U}(\Phi_0)=
		\{\Phi\;\vert\;\Phi = \Phi_0+U^{[iv,2]\bot}_{\Phi_0}[\mathbf{\gamma}_1\;\mathbf{\gamma}_2\;\cdots
		\;\mathbf{\gamma}_{m_z}]\}
	\end{equation}
in which for each $j=1,\cdots,m_z$, $\mathbf{\gamma}_j\in\mathcal{R}^{\widehat{r}(\Phi_0)}$. Obviously, this set includes the SCM $\Phi_0$ as an element. In addition, the proof of the above theorem reveals that any SCM $\Phi_u\in\mathscr{U}(\Phi_0)$ leads to an NDS TFM that is completely the same as that resulted from the SCM $\Phi_0$. According to Lemma \ref{lemma2}, this means that the associated NDS external responses $y(t,\Phi_u)$ and $y(t,\Phi_0)$ are identically equal to each other, provided these two NDS are stimulated by the same admissible external input signal. Hence, the NDS $\Sigma(\Phi_u)$ is also well-posed. In addition, if the NDS $\Sigma(\Phi_0)$ is stable, then the NDS $\Sigma(\Phi_u)$ is also stable.
\end{remark}

The set $\mathscr{U}(\Phi_0)$ is called the undifferentiable region associated with the SCM $\Phi_0$. Clearly, this set degenerates into the single element $\Phi_0$ when the condition of Theorem \ref{theo2} is satisfied. Otherwise, it is not bounded. In addition, Eq.(\ref{unidarea}) also indicates that global identifiability of the NDS structure at the specific SCM $\Sigma(\Phi_0)$ is equivalent to its local identifiability.

Now, we investigate structure identifiability for the NDS $\rm\bf\Sigma$ under the following situation.
\begin{enumerate}
\setcounter{enumi}{2}
\item For each $i\in \{1,2,\cdots,N\}$, the TFM $G_{zu}(\lambda,i)$ is of FNRR, but there exists at least one integer $j\in \{1,2,\cdots,N\}$, such that the TFM $G_{yv}(\lambda,j)$ is not of FNCR.
\end{enumerate}

To achieve this objective, partition the unimodular MVP $V^{[iv]}_{yv}(\lambda,i)$ into two row blocks,
\begin{equation}
	V^{[iv]}_{yv}(\lambda,i)=col\left\{
			V_{yv}^{[iv,1]}(\lambda,i),\;  V_{yv}^{[iv,2]}(\lambda,i)
	\right\}
\end{equation}
in which the MVP $V_{yv}^{[iv,1]}(\lambda,i)$ has $r_{yv}^{[i]}$ rows. As $V_{yv}^{[iv]}(\lambda,i)$ is unimodular, it is obvious that both the MVPs $V_{yv}^{[iv,1]T}(\lambda,i)$ and $V_{yv}^{[iv,2]T}(\lambda,i)$ are of FCR at every $\lambda\in\mathcal{C}$. In addition, the columns of the MVP $V_{yv}^{[iv,2]T}(\lambda,i)$ form a base of the right null space of the TFM $G_{yv}(\lambda,i)$. For each $i=1,2,\cdots,N$, denote by $\widehat{r}_{yv}^{[i]}$  the dimension of the right null space of the TFM $G_{yv}(\lambda,i)$. Then the MVP $V_{yv}^{[iv,2]T}(\lambda,i)$ has $\widehat{r}_{yv}^{[i]}$ columns.

Note that $D_{zv}^{-1}(\lambda,i)V_{yv}^{[iv,2]T}(\lambda,i)$ is an  $m_{v_i}\times\widehat{r}^{[i]}_{yv}$ dimensional RFM. There exist MVPs  $\widehat{R}(\lambda,i), \widehat{Q}(\lambda,i)\in \mathcal{R}^{m_{v_i}\times\widehat{r}^{[i]}_{yv}}[\lambda]$ and  $\widehat{\Omega}(\lambda,i)\in \mathcal{R}^{\widehat{r}^{[i]}_{yv}\times\widehat{r}^{[i]}_{yv}}[\lambda]$, such that $D^{-1}_{zv}(\lambda,i)V^{[iv,2]T}_{yv}(\lambda,i) =\widehat{R}(\lambda,i)+\widehat{Q}(\lambda,i)
\widehat{\Omega}^{-1}(\lambda,i)$ and the MVPs $\widehat{Q}(\lambda,i)$ and $\widehat{\Omega}(\lambda,i)$ are right coprime. Moreover, $\widehat{Q}(\lambda,i)\widehat{\Omega}^{-1}(\lambda,i)$ is strictly rational.

Define an integer $\widehat{r}_{yv}$, MVPs $\widehat{X}(\lambda)\in \mathcal{R}^{m_{v}\times\widehat{r}_{yv}}[\lambda]$ and $\widehat{Y}(\lambda) \in \mathcal{R}^{m_{z}\times \widehat{r}_{yv}}[\lambda]$ respectively as $\widehat{r}_{yv}=\sum^N_{i=1}\widehat{r}^{[i]}_{yv}$, and
\begin{align} \hspace*{-0.25cm}\widehat{X}(\lambda)&\! =\! diag\{D_{zv}(\lambda,i)[\widehat{R}(\lambda,i)\widehat{\Omega}(\lambda,i) \!+\! \widehat{Q}(\lambda,i)]|^N_{i=1}\}
	\\ \hspace*{-0.25cm}\widehat{Y}(\lambda)& \!=\! diag\{N_{zv}(\lambda,i)[\widehat{R}(\lambda,i)\widehat{\Omega}(\lambda,i) \!+\! \widehat{Q}(\lambda,i)]|^N_{i=1}\}
\end{align}
Moreover, let $\widehat{r}(\Phi_0)$ denote the normal rank of the MVP $\widehat{X}(\lambda)-\Phi_0\widehat{Y}(\lambda)$. Furthermore, assume that this MVP has the following Smith form,
\begin{equation}\label{Smith2}
\widehat{X}(\lambda)-\Phi_0\widehat{Y}(\lambda)=\widehat{U}_{\Phi_0}(\lambda)
	\left[\begin{array}{cc} diag\{\widehat{\mu}^{[j]}_{\Phi_0}(\lambda)|^{\widehat{r}(\Phi_0)}_{j=1}\} & 0  \\
0 & 0
\end{array}\right]
	\widehat{V}^T_{\Phi_0}(\lambda)
\end{equation}
Denote the inverse of the unimodular MVP $\widehat{U}_{\Phi_0}(\lambda)$ by $\widehat{U}^{[iv]}_{\Phi_0}(\lambda)$,  and partition it into two row blocks as follows,
\begin{equation}
	\widehat{U}^{[iv]}_{\Phi_0}(\lambda) = col \left\{
		\widehat{U}^{[iv,1]}_{\Phi_0}(\lambda), \;\; \widehat{U}^{[iv,2]}_{\Phi_0}(\lambda)
	\right\}
\end{equation}
in which $\widehat{U}^{[iv,1]}_{\Phi_0}(\lambda)$ has $\widehat{r}(\Phi_0)$ rows. Define an integer $\widehat{p}(\Phi_0)$ as $\widehat{p}(\Phi_0) = \max_{i,j}\{deg[\widehat{U}^{[iv,2]}_{\Phi_0}(\lambda)]_{ij}\}$. Then there exists a set and only a set of real matrices $\widehat{U}_{\Phi_0,k}^{[iv,2]}$ with $0\leq k \leq \widehat{p}(\Phi_0)$, such that   $\widehat{U}^{[iv,2]T}_{\Phi_0}(\lambda)=\sum^{\widehat{p}(\Phi_0)}_{k=0}\lambda^k
\widehat{U}_{\Phi_0,k}^{[iv,2]}$.

\begin{corollary}\label{coro0}
Assume that the NDS $\rm\bf\Sigma$ satisfies simultaneously Assumptions A1) and A3). Then the structure of this NDS is globally identifiable at the SCM $\Phi_0$, if and only if the matrix $\widehat{U}_{\Phi_0}^{[iv,2]} =col\{\widehat{U}_{\Phi_0,k}^{[iv,2]}|^{\widehat{p}(\Phi_0)}_{k=0}\}$ is of FCR.
\end{corollary}

\noindent\textbf{Proof:} Under the assumption that the NDS $\rm\bf\Sigma$ itself and each of its subsystems are regular, as well as that all the TFMs $G_{zu}(\lambda)|_{i=1}^{N}$ are of FNRR, it can be directly claimed from Lemma \ref{lemma2} and Eq.(\ref{a.11}) that the structure of the NDS $\rm\bf\Sigma$ is globally identifiable at the SCM $\Phi_{0}$, if and only if there exists a nonzero $\Delta_\star\in \mathcal{R}^{m_{v}\times m_{z}}$, such that
\begin{equation}
G_{yv}(\lambda)[I_{m_v}-\Phi_0G_{zv}(\lambda)]^{-1}\Delta_\star
\equiv  0
\label{eqn:3-1}
\end{equation}

On the basis of this equality, arguments similar to the proofs of Theorems \ref{theo1} and \ref{theo2} lead to the conclusions.
This completes the proof.   \hspace{\fill}$\Diamond$

As in Theorem \ref{theo2}, when the matrix $\widehat{U}_{\Phi_0}^{[iv,2]}$ is not of FCR, its right null space defines the undifferentiable region associated with the SCM $\Phi_0$. In addition, the nonuniqueness of the right MFD $\widehat{Q}(\lambda,i)\widehat{\Omega}^{-1}(\lambda,i)$ with $1\leq i\leq N$, does not affect the conclusions of Corollary \ref{coro0}.

Note that for arbitrary SCMs $\Phi_{1},\Phi_{2}\in \mathscr{S}$, $H(\lambda,\Phi_{1}) \equiv H(\lambda,\Phi_{2})$ if and only if
$H^{T}(\lambda,\Phi_{1}) \equiv H^{T}(\lambda,\Phi_{2})$. It can therefore be declared from Eq.(\ref{tfm}) and Lemma \ref{lemma2} that Corollary \ref{coro0} can be directly applied to the situation in which there is a $1\leq i\leq N$ such that the TFM $G_{zu}(\lambda,i)$ is not of FNRR, while the TFM $G_{yv}(\lambda,j)$ is of FNCR for each $1\leq j\leq N$. When each subsystem is described by a state space model, results similar to Corollary \ref{coro0} have been obtained in \cite{yz2021}.

When for each $1\leq j\leq N$, the TFMs $G_{zu}(\lambda,i)$ and $G_{yv}(\lambda,i)$ are respectively of FNRR and FNCR, it has been proven in \cite{Zhou2020-1} that the structure of the NDS $\rm\bf\Sigma$ is globally identifiable, provided that the dynamics of each subsystem is described by a state space model, meaning that $E(i) = I_{m_{xi}}$ for each $i=1,2,\cdots,N$. These conclusions remain valid for the NDS adopted in this paper, and the proof is completely the same. The details are therefore omitted.

\section{Some Particular Situations}

In actual applications, there are usually some a priori information about NDS subsystem interactions. Two situations are investigated in this section about NDS structure identifiability that are often encountered in practice. One is that some NDS subsystem interactions are known exactly, and the other is that the unknown subsystem interactions are not algebraically independent of each other.

In a real world NDS, some subsystem interactions may be known exactly from its working principles, while some subsystem connections are forbidden or not possible from its constructions \cite{htw2009,pmssaxcas2010,Siljak1978,zyl2018}. An obvious example is that self-loops are usually not considered in NDS analysis and synthesis, meaning that the diagonal blocks of the SCM $\Phi$ are usually fixed to be a zero matrix. Generally, this kind of NDS structure information can be reflected by defining the set $\mathscr{S}$ as follow,
\begin{equation}\label{sset}
	\mathscr{S}=\{\; \Phi \;|\; [\Phi]_{ij}\; \text{is known},\; i\in\mathbb{I}_j,\; j\in\mathbb{J}\; \}
\end{equation}
in which
$\mathbb{I}_j=\{k_{j,1},k_{j,2},\cdots,k_{j,q(j)}\}$ with $1\le k_{j,1}<\cdots<k_{j,q(j)}\le m_v$ and $q(j)\le m_v$, and $\mathbb{J}\subseteq\{1,\cdots,m_z\}$.
That is, $\mathbb{I}_j \subseteq \{1,\cdots,m_v\}$ contains
$q(j)$ elements indicating positions of the elements in the $j$-th column of the SCM $\Phi$ whose values are exactly known.

Concerning the matrix $U^{[iv,2]}_{\Phi_0}$ defined in Theorem \ref{theo2}, denote its $i$-th column vector by ${u}^{[iv,2]}_{\Phi_0,i}$ with $i=1,\cdots,m_v$. Using these symbols, the following conclusions are obtained.

\begin{corollary}\label{coro1}
Assume that the a priori structure information of the NDS $\rm\bf\Sigma$ is described by Eq.(\ref{sset}). Moreover, assume that the NDS $\rm\bf\Sigma$ satisfies both Assumptions A1) and A2). Then the structure of the NDS $\Sigma$ is globally identifiable at a specific SCM $\Phi_0\in \mathscr{S}$, if and only if each $j\in\mathbb{J}$, the matrix $U^{[iv,2]}_{\Phi_0,j}$ defined as follow is of FCR,
	\begin{equation}\label{corollary1}
		U^{[iv,2]}_{\Phi_0,j} = \{\; {u}^{[iv,2]}_{\Phi_0,i} \;|\; i\in\{1,\cdots,m_v\}\setminus\mathbb{I}_j\}
	\end{equation}
\end{corollary}

\noindent\textbf{Proof:} Let $\Phi_{\star}$ be an SCM distinctive from $\Phi_0$, and denote $\Phi_0-\Phi_\star$ by $\Delta_\star$. Then $\Phi_{\star}\in \mathscr{S}$ if and only if $[\Delta_\star]_{ij}=0$ whenever $i\in\mathbb{I}_j$ and $j\in\mathbb{J}$. The proof can now be completed through putting these restrictions on the vector $\delta$ of Eq.(\ref{a.23}).   \hspace{\fill}$\Diamond$

\begin{remark}
Assume that there is an NDS $\rm\bf\Sigma$ whose
a priori structure information is described by the set $\mathscr{S}$ of Eq.(\ref{sset}), and the conditions of Corollary $\ref{coro1}$ are not satisfied by this NDS at a specific SCM $\Phi_0\in \mathscr{S}$. These mean that there is at least one $j\in\mathbb{J}$ such that the right null space of the matrix $U^{[iv,2]}_{\Phi_0,j}$ is not trivial. For each $j\in\mathbb{J}$, let $\widehat{r}(\Phi_0,j)$ stand  for the dimension of the right null space of the matrix $U^{[iv,2]}_{\Phi_0,j}$. Then the undifferentiable region $\mathscr{U}_s(\Phi_0)$ of the NDS $\rm\bf\Sigma$ associated with this particular SCM $\Phi_0\in \mathscr{S}$, can be represented as
\begin{equation}
		\mathscr{U}_s(\Phi_0)=	\{\;\Phi\;\vert\;\Phi=\Phi_0 + \Delta\;\}
\end{equation}
in which $\Delta\in \mathcal{R}^{m_{v}\times m_{z}}$, and for each $i\in\{1,2,\cdots,m_v\}$ and each $j\in\{1,2,\cdots,m_z\}$,  its $i$-th row $j$-th column element $[\Delta]_{ij}$ is defined as
\begin{equation*}
		\Delta_{ij} \stackrel{\Delta}{=}\biggr\{
		\begin{array}{cl}
			0, & i\in\mathbb{I}_j\\
			e_{s}^{T}U^{[iv,2]\bot}_{\Phi_0,j}\gamma_{j}, & s=i-|\{1,\cdots,i\}\cap\mathbb{I}_j|, \quad  i\notin\mathbb{I}_j
		\end{array}
\end{equation*}
Here, $e_{j}$ is the $j$-th standard basis of the Euclidean space $\mathcal{R}^{m_{v}}$, while $\gamma_{j}\in \mathcal{R}^{\widehat{r}(\Phi_0,j)}$ is an arbitrary real vector.
\end{remark}

Note that in Eq.(\ref{sset}), the unknown elements of the SCM $\Phi$, that is, $[\Phi]_{ij}$ with $i\in\{1,\cdots,m_v\} \backslash \mathbb{I}_j$ when $j\in \mathbb{J}$ and $[\Phi]_{ij}$ with $i\in\{1,\cdots,m_v\}$ when $j\in \{1,2,\cdots,m_z\} \backslash\mathbb{J}$, are implicitly assumed to be mutually algebraically  independent. This may not always be satisfied in actual applications, noting that subsystems in many real world NDSs are connected through some physical elements, etc., which usually makes their  subsystems share some parameters \cite{Siljak1978,zyl2018}. In order to take these algebraic dependence among subsystem interactions into account, it can be generally assumed that the SCM $\Phi$ of the NDS $\rm\bf\Sigma$ meets the following constraints
\begin{equation}\label{Phik}
	\Phi(\theta)=\Phi^{[0]}+\sum^{q}_{k=1}\theta^{[k]}\Phi^{[k]}
\end{equation}
in which $\theta$ denotes the vector $col\{\theta^{[k]}|_{k=1}^{q}\}$, $\Phi^{[0]} \in \mathcal{R}^{m_v\times m_z}$ is a fixed constant matrix representing a priori information on NDS structure, $\Phi^{[k]} \in \mathcal{R}^{m_v\times m_z}$ with $k=1,2,\cdots,q$, is also a fixed constant matrix reflecting some known structure information about algebraic dependence among subsystem interactions, while $\theta^{[k]}\in\mathbb{R}$ with $k=1,2,\cdots,q,$ is an unknown parameter to be estimated from experiment data, etc., and $q$ is the number of algebraically independent parameters of the SCM $\Phi$.

Eq.(\ref{Phik}) reveals that some elements of the NDS SCM $\Phi$ are algebraically dependent, leading to the following conclusions about NDS structure identifiability.

\begin{corollary}\label{coro2}
Assume that the NDS $\rm\bf\Sigma$ satisfies Assumptions A1) and A2),  and its SCM $\Phi$ is constrained by Eq.(\ref{Phik}). Then the NDS structure is globally identifiable at a particular parameter vector $\theta_0=col\{\theta_0^{[k]}|_{k=1}^{q}\}$, if and only if the following matrix is of FCR,
\begin{equation}
\left[vec(U^{[iv,2]}_{\Phi(\theta_0)}\Phi^{[1]})\quad vec(U^{[iv,2]}_{\Phi(\theta_0)}\Phi^{[2]})\; \cdots \; vec(U^{[iv,2]}_{\Phi(\theta_0)}\Phi^{[q]})\right]
\end{equation}
\end{corollary}

\noindent\textbf{Proof:} Let $\Phi(\theta_{\star})$ be an SCM satisfying Eq.(\ref{Phik}) in which $\theta_{\star}=col\{\theta_{\star}^{[k]}|_{k=1}^{q}\}$. Denote $\Phi(\theta_{\star})-\Phi(\theta_{0})$ by $\Delta_\star$. Then
\begin{equation}\label{Phik-1}
\Delta_\star = \sum^{q}_{k=1}\left(\theta_{\star}^{[k]} - \theta_{0}^{[k]}\right)\Phi^{[k]}
\end{equation}
The proof can now be completed by substituting each column vector of the above matrix $\Delta_\star$ into Eq.(\ref{a.23}).   \hspace{\fill}$\Diamond$

From the constraint descriptions, it is clear that an SCM of the set $\mathscr{S}$ of Eq.(\ref{sset}) can also be expressed as that of Eq.(\ref{Phik}). This means that Corollary $\ref{coro2}$ is also applicable to the situation dealt by Corollary $\ref{coro1}$.  Nonetheless, computational complexity for verifying the condition of Corollary $\ref{coro1}$ is in general far less than that of Corollary $\ref{coro2}$. On the other hand, when the condition of Corollary $\ref{coro2}$ is not satisfied, an explicit expression can also be obtained for the associated undifferentiable region. The details are omitted due to its obviousness and space considerations.

In many practical applications, rather than a precise parameter value, a priori information about the NDS structure includes only the positions of the SCM nonzero elements. For instance, on the basis of some biological knowledge about a gene regulation network, it is usually much easier to infer genes or other biochemical molecules that directly affect a specific gene, but rather hard and/or too expensive to infer an interaction strength \cite{htw2009,pmssaxcas2010,zyl2018}. On the other hand, all the aforementioned conditions for NDS structure identifiability require the Smith form of a MVP, which is computationally demanding for a large scale NDS. For example, the Smith form of the MVP $X(\lambda)-\Phi_0 Y(\lambda)$ must be computed in Theorem $\ref{theo2}$, whose computational complexity increases in general cubically with the MVP dimension. Thus when the NDS has a great number of subsystems, it may be  quite hard to actually calculate this form and therefore to verify the associated conditions. To overcome these difficulties, the following corollary is derived from Theorem $\ref{theo1}$.

\begin{corollary}\label{coro3}
Assume that the NDS  $\rm\bf\Sigma$ simultaneously satisfies Assumptions A1) and A2). Let $P \in \mathcal{R}^{m_z \times m_z}$ be a diagonal matrix with its diagonal elements algebraically independent and not identically equal to zero. Then the structure of the NDS  $\rm\bf\Sigma$ is globally identifiable at a specific SCM $\Phi_0$, if and only if for any nonzero vector $\delta\in\mathcal{R}^{m_v}$, the following equation does not have  a solution, in which $\alpha(\lambda)$ and $\beta(\lambda)$ are VVPs that may have an infinite degree,
\begin{equation}
		\begin{bmatrix}
			X(\lambda) & -\Phi_0 P \\ Y(\lambda) & -P
		\end{bmatrix}
		\begin{bmatrix}
			\alpha(\lambda) \\ \beta(\lambda)
		\end{bmatrix}=
		\begin{bmatrix}
			\delta \\ 0
		\end{bmatrix}
\label{eqn:coro3-1}
\end{equation}
\end{corollary}

\noindent\textbf{Proof:} Denote $P^{-1}Y(\lambda)\alpha(\lambda)$ by $\beta(\lambda)$. Then it can be understood without significant difficulties that Eq.(\ref{lemma}) and Eq.(\ref{eqn:coro3-1}) are equivalent to each other. This completes the proof.   \hspace{\fill}$\Diamond$

When the nonzero elements of the SCM $\Phi_0$ are algebraically independent, straightforward algebraic manipulations show that all the nonzero elements in the matrices $\Phi_0 P$ and $P$ also have this property. Then from the above corollary and relevant conclusions about a  valuated matroid \cite{KM2010}, a graph based verification procedure can be developed for global NDS structure identifiability. In this procedure, the calculation of a Smith form is no longer required. In addition, rather than the concrete values of the nonzero elements in the SCM $\Phi_0$, it is the positions of these nonzero elements that are needed. However, when these nonzero elements are not algebraically independent, further efforts are still required. Details are omitted for space considerations.

When for each $i\in\{1,2,\cdots,N\}$, the TFM $G_{zu}(\lambda,i)$ is of FNRR, or the TFM $G_{yv}(\lambda,i)$ is of FNCR, similar results can be obtained from Corollary \ref{coro0}.

Compared with the results of \cite{ptt2019,vtc2021}, the conditions of Sections 3 and 4 are both necessary and sufficient. In addition, they do not ask that the internal output vector of each subsystem is identically equal to its external one. On the other hand, while some similar results have been derived in \cite{yz2021,Zhou2020-1}, the subsystem dynamics are described by a state space model there, and some stringent constraints are to be satisfied by the TFMs $G_{zu}(\lambda,i)$, $G_{yv}(\lambda,i)$ and $G_{zv}(\lambda,i)$ for each $i\in\{1,2,\cdots,N\}$. It appears that the results of these two sections are the most general from both the adopted model and the requirements.

\section{SCM Reconstruction}

In actual applications, in addition to verify whether or not the structure of an NDS is identifiable, it is usually also very important to reveal its structure. While it may be the best way to estimate an NDS  structure directly from experiment data, some essential issues are still to be settled, including probing signal designs, etc. Another approach to estimate subsystem interactions is to divide the task into two steps \cite{sd2020,vtc2021}. At first, a frequency/time domain model of the overall NDS is estimated from experiment data, for which a method developed for multi-variable system identification can in principle be utilized. The latter has been extensively studied for a long time and many relatively matured procedures are already available \cite{Ljung1999,wvd2018}. The second step is to recover the subsystem interactions from the estimated NDS model, for which further efforts are still required.

In this section, we study how to reconstruct the SCM $\Phi$ of the NDS $\rm\bf\Sigma$ from its descriptor form model. For this purpose, a descriptor form model of the NDS $\rm\bf\Sigma$ is denoted by ${\rm\bf\Sigma}(A:E)$ which is in the form of Eq.(\ref{D}).

For a composite system to work properly that is constituted from several subsystems, it is important that the composite system is well-posed, which essentially means that the model of the whole composite system is well defined \cite{Kailath1980,Siljak1978,zdg1996,zyl2018}. Clearly, this is also a prerequisite for NDS subsystem interaction estimations. From \cite{Zhou2020-2}, we have that the NDS ${\rm\bf \Sigma}$ adopted in this paper, that is, the NDS described by Eqs.(\ref{dyn}) and (\ref{scm}), is well-posed, if and only if the matrix $I_{m_v}-\Phi D_{zv}$ is invertible. Substituting  Eq.(\ref{scm}) into Eq.(\ref{dyn}), it is not quiet difficult to understand that this requirement is essential to guarantee that at each time instant, the NDS internal output vector $z(t)$ is uniquely determined by its state vector $x(t)$ and external input vector $u(t)$.

When the NDS $\rm\bf\Sigma$ is well-posed, straightforward algebraic manipulations show that, the dynamics of the whole NDS $\rm\bf\Sigma$ can be described as follows,
\begin{multline}\label{SSM}
	\begin{bmatrix}
		E\delta(x(t)) \\ y(t)
	\end{bmatrix}=\Biggr\{
	\begin{bmatrix}
		A_{xx} & B_{xu} \\
		C_{yx} & D_{yu}
	\end{bmatrix}+
	\begin{bmatrix}
		B_{xv} \\ D_{yv}
	\end{bmatrix}
    (I_{m_v}-\Phi D_{zv})^{-1}\\
    \times\Phi
	\begin{bmatrix}
		C_{zx} & D_{zu}
	\end{bmatrix}\Biggr\}
	\begin{bmatrix}
		x(t) \\ u(t)
	\end{bmatrix}
\end{multline}
which also takes completely the same form of a descriptor system given by Eq.(\ref{D}).

On the basis of this equation, the following results are obtained. Their  proof is deferred to the appendix.

\begin{theorem}\label{theo4}
Assume that the NDS is well-posed. Then the SCM $\Phi$ can be reconstructed from an consistent descriptor form model of the whole NDS, if and only if the matrix $col\{B_{xv},\; D_{yv}\}$ is of FCR, and the matrix $[C_{zx} \;\; D_{zu}]$ is of FRR.
\end{theorem}

Compared with the results of \cite{sd2020} which are based on the TFM of the whole NDS, the conditions of Theorem \ref{theo4} can be verified with each subsystem independently, noting that both the matrix $col\{B_{xv},\; D_{yv}\}$ and the matrix $[C_{zx} \;\; D_{zu}]$ are block diagonal, indicating that these conditions are equivalent to that for each $i=1,2,\cdots,N$, the matrices  $col\{B_{xv}(i),\; D_{yv}(i)\}$ and $[C_{zx}(i) \;\; D_{zu}(i)]$ are respectively of FCR and of FRR, which can be separately checked for each subsystem independently. In other words, the associated computational complexity increases linearly with the NDS subsystem number, which makes these conclusions attractive in the analysis and synthesis of a large-scale NDS. On the other hand, the NDS model adopted here is more general than the structured linear fractional transformation model of \cite{sd2020}.

Another attractive characteristic of the conditions in Theorem \ref{theo4} is that they do not depend on a particular estimated NDS model, meaning that structure reconstructibility is also a property held by an NDS model and has no concerns with experiment data, estimation procedures, etc.

Note also that the matrix  $col\{B_{xv}(i),\; D_{yv}(i)\}$ reflects influences of the internal input vector $v(t,i)$ of the $i$-th subsystem ${\rm\bf\Sigma}_{i}$ on its state vector $x(t,i)$ and its external output vector $y(t,i)$, while the matrix $[C_{zx}(i) \;\; D_{zu}(i)]$ influences from its state vector $x(t,i)$ and its external input vector $u(t,i)$ to  its internal output vector $z(t,i)$. The conditions of Theorem \ref{theo4} also answer to some extent an important question in NDS designs. That is, in order to make the structure of an NDS identifiable, what kind of signals must be sent to other subsystems, and what kind of influences from other subsystems can be accepted, recalling that an NDS TFM may have several descriptor form realizations \cite{Dai1989,Kailath1980,zdg1996,zyl2018}, while NDS structure identifiability requires that its SCM is solely determined by its TFM which is clarified by Lemma \ref{lemma2}.

On the other hand, from Eq.(\ref{SSM2}), the following results are obtained for the consistency of a descriptor form model of the NDS $\rm\bf\Sigma$ with its structure.

\begin{corollary}\label{coro4}
Assume that the NDS is well-posed and its subsystem interactions are reconstructible from its descriptor form model. Then a descriptor form model ${\rm\bf\Sigma}(\widehat{A}:\widehat{E})$ of the whole NDS ${\rm\bf\Sigma}$ is consistent with the structure of the NDS ${\rm\bf\Sigma}$, if and only if the following three conditions are satisfied simultaneously
\begin{equation}
\hspace*{-0.5cm} K^{\bot}_{l}E_{d} =0,
\hspace{0.1cm} E_{d}L^{\bot}_{r} =0,
\hspace{0.1cm} \left\{I_{m_v} \!+\! H_{m} D_{zv}\right\}_{l}^{\bot} H_{m} =0
\label{eqn:c4-1}
\end{equation}
in which
\begin{eqnarray*}
& & K=col\{B_{xv},\; D_{yv}\}, \hspace{0.15cm} H_{m}=(K^{T}K)^{-1}K^{T}E_{d}L^{T}(LL^{T})^{-1} \\
& & L=[C_{zx} \;\; D_{zu}],\hspace{0.5cm}
E_{d}= \begin{bmatrix}
		\widehat{A} & \widehat{B} \\ \widehat{C} & \widehat{D}
	\end{bmatrix}
\!-\!
	\begin{bmatrix}
	A_{xx} & B_{xu} \\
	C_{yx} & D_{yu}
	\end{bmatrix}
\end{eqnarray*}
\end{corollary}

From the proofs of Theorem \ref{theo4} and Corollary \ref{coro4}, it is clear that if the subsystem interactions of the NDS ${\rm\bf\Sigma}$ are reconstructible, and a descriptor form model ${\rm\bf\Sigma}(\widehat{A}:\widehat{E})$ is consistent, then its SCM $\Phi$ is uniquely determined by
\begin{equation}
\Phi = \left(I_{m_v}+ H_{m} D_{zv}\right)^{-1} H_{m}
\label{eqn:scm-recover}
\end{equation}
which is obviously a continuous function of the system matrices in the descriptor form model, meaning that estimation errors on the whole NDS model usually have a bounded influences on the recovery of its SCM $\Phi$.

\begin{figure}[!htb]
	%\vspace{-1cm}
	%	\setlength{\belowcaptionskip}{0.1cm}
	\centering
	\begin{minipage}{\columnwidth}
		\hspace*{0.0cm}\includegraphics[width=\linewidth]{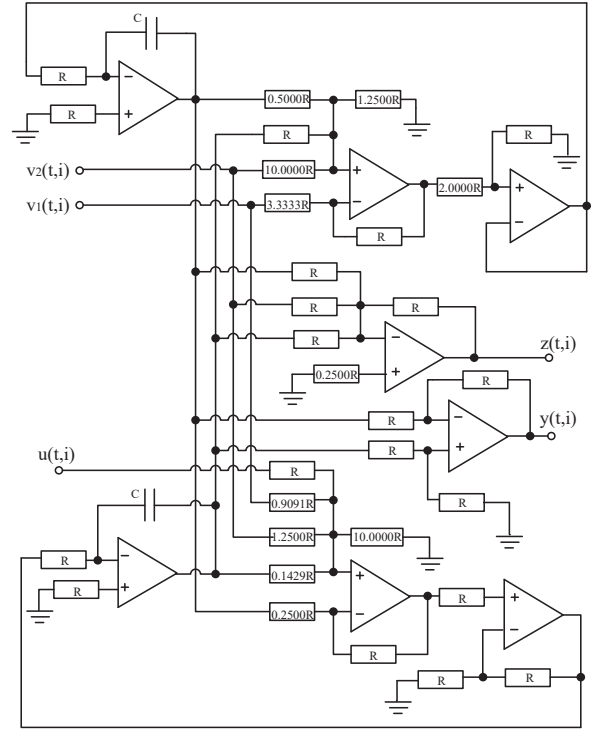}
	\end{minipage}
\vspace{-0.5cm}
	\caption{A subsystem of the artificial NDS.}
\label{Fig-x}
\end{figure}

When an NDS model is given in the frequency domain, meaning that an estimate of its TFM is available, further efforts are still required on reconstructing its SCM $\Phi$.

\section{A Numerical Example}

In order to illustrate the obtained conditions for NDS structure identifiability, an artificial NDS is constructed in this section. This NDS is constituted from two subsystems, and each subsystem consists of 2 capacitors, 8 operation amplifiers, and 31 resistors, as shown in Fig.\ref{Fig-x}. To have a concise presentation without sacrificing engineering significance, all the resistors are set to a value  proportional to $R$ which are given in the aforementioned figure. In addition, all the capacitors are set to the same value $C=1.0000/R$. Using these physical parameter values, system matrices of Subsystem $\Sigma_i$ with  $i=1,2$, are directly obtained from circuit principles, which are given as follows,
\begin{eqnarray*}
& & A_{xx}(i) \!=\!
		\begin{bmatrix}\!
			-2.0000 & -1.0000 \\ 4.0000 &-7.0000
		\!\end{bmatrix}\!\!,\hspace{0.20cm}
    B_{xv}(i) \!=\!
		\begin{bmatrix}\!
			-0.3000 & 0.1000 \\ 1.1000 & 0.8000
		\!\end{bmatrix} \\
& & B_{xu}(i)=
		\begin{bmatrix}
			0 \;\; 1.0000
		\end{bmatrix}^{T}\!,\hspace{0.25cm} C_{zx}(i)=
		\begin{bmatrix}
			1.0000 & 1.0000
		\end{bmatrix}  \\
& & C_{yx}(i)=
		\begin{bmatrix}
			1.0000 & -1.0000
		\end{bmatrix} \!,\hspace{0.25cm}
    D_{zv}(i)=
	    \begin{bmatrix}
	    	0 & -1.0000
	    \end{bmatrix}   \\		
& & D_{yv}(i)=
	    \begin{bmatrix}
	    	0 & 0	    \end{bmatrix}\!,\hspace{0.25cm}D_{yu}(i)=D_{zu}(i)=0,\hspace{0.25cm}
E(i)=I_{2}
\end{eqnarray*}

To investigate time domain differences between two distinctive SCM associated NDSs, say $\Sigma(\Phi_1)$ and $\Sigma(\Phi_2)$, the following settings have been adopted in numerical simulations. Denote by $A_1$ and $A_2$ the state transition matrices (STM) of the NDSs $\Sigma(\Phi_1)$ and $\Sigma(\Phi_2)$ respectively. The sampling period $T$ is set to  $\frac{0.1000}{max\{\rho_{max}(A_1),\rho_{max}(A_2)\}}$, while the  sampling number $M$ to $max\left\{10^4,\left\lfloor 100\times\frac{max\{\rho_{max}(A_1),\rho_{max}(A_2)\}}{min\{\rho_{min}(A_1),\rho_{min}(A_2)\}}\right\rfloor\right\}$. In addition, two independent pseudo-random binary signals (PRBS) are utilized as the external inputs to the NDS $\rm\bf\Sigma$, which take values from the set $\{-10.0000,\; 10.0000\}$. Each PRBS signal is inputted to one subsystem respectively. These settings are expected to guarantee that the input signals are exciting and informative enough to stimulate the differences between the outputs of two different NDSs, but do not bring a high computational cost.

It is worthwhile to mention that \cite{wrmd2005} has illuminated that a persistently exciting input signal, together with its associated output signal, is able to represent all the input/output behaviors of any finite dimensional LTI dynamic system, no matter what the input signal is. This means that two different SCMs $\Phi_{1}$ and $\Phi_{2}$ in the set $\mathscr{S}$ are not differentiable from each other, if and only if ${\rm\bf\Sigma}(\Phi_{1})$ and ${\rm\bf\Sigma}(\Phi_{2})$ respond identically to a persistently exciting external input signal. Note also that a PRBS is widely known to have this property and has been extensively adopted in system identification \cite{Ljung1999,wvd2018}. It therefore appears safe to declare that the aforementioned numerical experiment settings are reasonable.

At first, the following three SCMs $\Phi_0$, $\Phi_u$ and $\Phi_i$ are considered. It can be directly verified that each of them leads to a regular and stable NDS, and the associated NDS is also well-posed.
\begin{small}
	\setlength{\abovedisplayskip}{0.01pt}
	\setlength{\belowdisplayskip}{0.01pt}
	\begin{gather*}
		\Phi_0=
		\begin{bmatrix}
			0 & 0 \\
			0 & 0 \\
			1.0000 & 0 \\
			0 & 0
		\end{bmatrix},
		\Phi_u=
		\begin{bmatrix}
			0 & 0 \\
			0 & 0 \\
			0 & 0 \\
			2.0000 & 0
		\end{bmatrix},
			\Phi_i=
		\begin{bmatrix}
			0 & 1.0000 \\
			0 & 0 \\
			1.0000 & 0 \\
			0 & 0
		\end{bmatrix}
	\end{gather*}
\end{small}

On the basis of the aforementioned system matrices for each subsystem, direct computations show that the TFM $G_{zu}(\lambda)$ is of FNRR, while the TFM $G_{yv}(\lambda)$ is not of FNCR. This means that Corollary \ref{coro0} is relevant to this example. In addition, the matrix $\widehat{U}^{[iv,2]}_{\Phi_0}$ of Corollary \ref{coro0} takes the following value and is obviously not of FCR.
\begin{small}
\setlength{\abovedisplayskip}{0.01pt}
\setlength{\belowdisplayskip}{0.01pt}
	\begin{equation*}
		\widehat{U}^{[iv,2]}_{\Phi_0} =
		\begin{bmatrix}
			-5.0000&0&-3.0000&-1.5000\\
			2.0000&1.0000&0&0\\
			-2.0000&0&-1.0000&-0.5000\\
			0&0&0&0
		\end{bmatrix}
	\end{equation*}
\end{small}

According to Corollary \ref{coro0}, it can be claimed that the structure of this artificial NDS $\Sigma$ is not globally identifiable at this specific SCM $\Phi_0$. That is, there is at least one SCM $\Phi \neq{\Phi_0}$ resulting in an NDS $\Sigma(\Phi)$ whose external output  $y(t,\Phi)$ can not be differentiated from that of $\Sigma(\Phi_0)$, no matter what external stimulations are used. All these SCMs are given by the set $\mathscr{U}(\Phi_0)$ of Eq.(\ref{unidarea}) with the matrix ${U}^{[iv,2]}_{\Phi_0}$ being replaced by the matrix $\widehat{U}^{[iv,2]}_{\Phi_0}$. From this description, it can be easily understood that $\Phi_{u}\in \mathscr{U}(\Phi_0)$, while $\Phi_{i}\not\in \mathscr{U}(\Phi_0)$.

\begin{figure}[!htb]
	\centering
		\begin{minipage}{0.45\columnwidth}
\includegraphics[width=\linewidth]{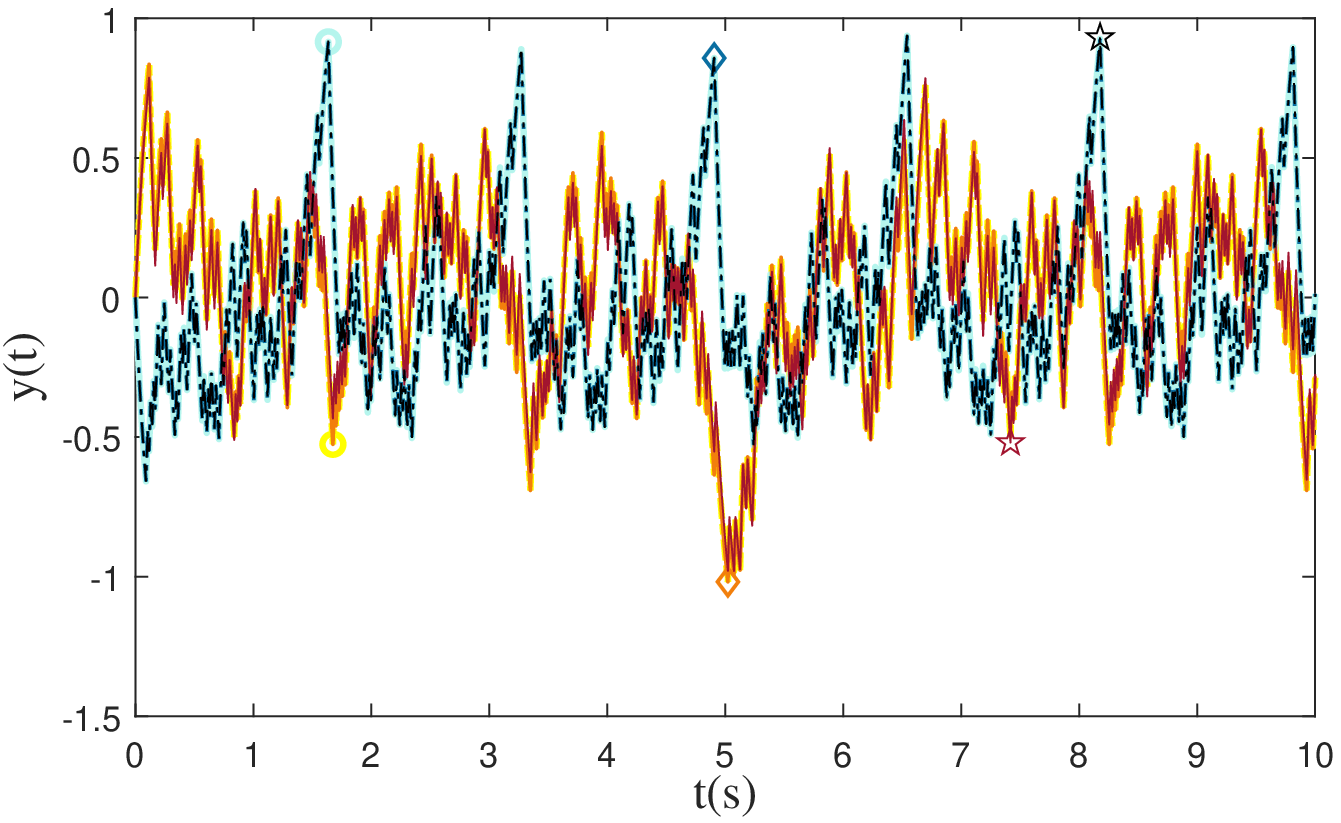}
{\small \hspace*{0.5cm} (a) external output.}
	\end{minipage}
\hspace{0.2cm}
		\begin{minipage}{0.45\columnwidth}
\includegraphics[width=\linewidth]{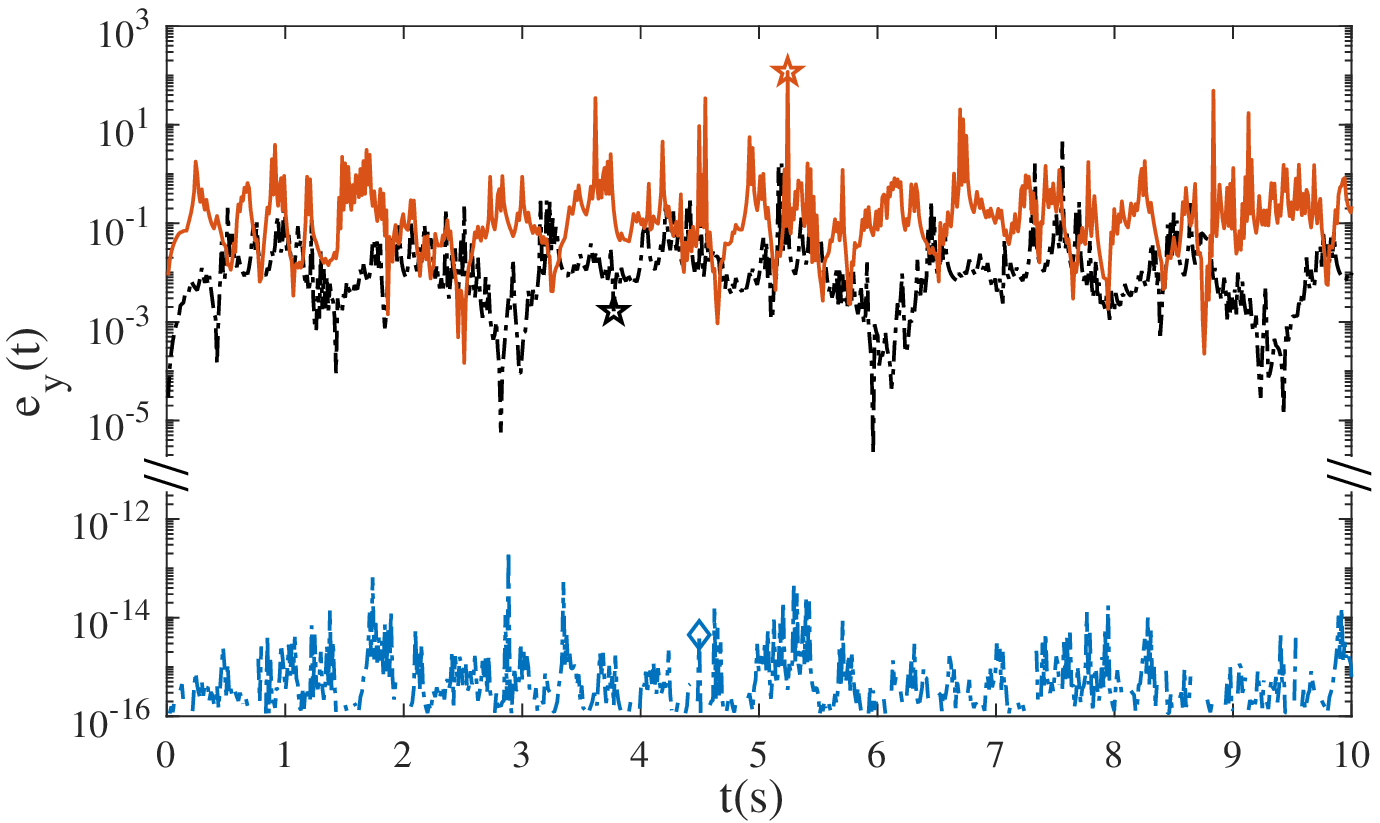}
{\small \hspace*{0.5cm}(b) relative differences. }
	\end{minipage}
\caption{NDS external outputs and their relative differences.
$\circ$: $\Phi_0$; $\star$: $\Phi_i$; $\Diamond$: $\Phi_u$. $-\!\!-$: the 1st subsystem; $-\cdot-$: the 2nd subsystem. }
\label{figure1}
\end{figure}

On the other hand, the matrices $K$ and $L$ of Corollary $\ref{coro4}$ are respectively as follows,
\begin{equation*}
	K=
	\begin{bmatrix}
	-0.3000 & 0.1000 \\ 1.1000 & 0.8000 \\ 0 & 0	
	\end{bmatrix},
	L=
	\begin{bmatrix}
	1.0000 & 1.0000 & 0
	\end{bmatrix}
\end{equation*}
Clearly, the matrix $K$ is of FCR and the matrix $L$ is of FRR. This implies that if a consistent descriptor form model is available for the whole NDS $\rm\bf\Sigma$, then its SCM can be uniquely determined. These observations confirm that NDS reconstructibility from a time domain model is not equivalent to its identifiability, and is in general only necessary for the latter.

To see the difference between the external outputs of the NDSs   $\Sigma(\Phi_0)$ and $\Sigma(\Phi_\star)$ with $\star = u$ or $i$, define a relative error $e_{y,\star}(t,j)$ for each subsystem pair as
\begin{displaymath}
e_{y,\star}(t,j)=\left|\frac{y(t,j,\Phi_\star)-y(t,j,\Phi_0)}{y(t,j,\Phi_0)}\right|,\hspace{0.5cm} j=1,2
\end{displaymath}

Fig.$\ref{figure1}$ shows the simulated external subsystem outputs of the NDSs $\Sigma(\Phi_0)$, $\Sigma(\Phi_u)$ and $\Sigma(\Phi_i)$, together with the corresponding relative differences. As the relative differences between the 1st subsystems of the NDSs $\Sigma(\Phi_0)$ and $\Sigma(\Phi_u)$ are identically equal to zero, they are not shown in Fig.\ref{figure1}b. From Fig.\ref{figure1}b, it is clear that although the relative difference $e_{y,u}(t,2)$ is not identically equal to zero, it is very close to zero, noting that its maximum value is in the order of $10^{-12}$, which can be viewed as effects of numerical calculation errors. On the contrary, Fig.\ref{figure1}b also shows that the maximum value of the relative difference $e_{y,i}(t,j)$ with $j\in \{1,2\}$ is about $100$, indicating that there do exist differences between the outputs of the NDS $\Sigma(\Phi_0)$ and the NDS $\Sigma(\Phi_i)$.

These observations agree well with Corollary \ref{coro0}, noting that the SCM $\Phi_{u}$ belongs to the set $\mathscr{U}(\Phi_0)$, which is the undifferentiable region of the SCM $\Phi_0$, while the SCM $\Phi_{i}$ does not.

Next, relations are investigated between the distance of an SCM $\Phi$ to the undifferentiable region $\mathscr{U}(\Phi_0)$ of the SCM $\Phi_{0}$ and the corresponding NDS output differences.

For this purpose, 4 random SCM samples $\widetilde{\Phi}_i|_{i=1}^{4}$ are at first generated through $\widetilde{\Phi}_i = {\Phi}_0 + \Delta_{i}$, in which each element of $\Delta_{i}|_{i=1}^{4}$  has an independent and continuous uniform distribution over the interval $[-2.0000, \;2.0000]$. The values of these SCMs are as follows,
\begin{small}
	\begin{align*}
		\widetilde{\Phi}_1&=
		\begin{bmatrix}			1.1732&0.4875\\-0.5079&-0.4236\\2.3282&-0.5629\\1.0153&-1.6446
		\end{bmatrix}\!,\hspace{0.25cm}
		\widetilde{\Phi}_2=
		\begin{bmatrix}
0.4634&-1.5646\\-1.3977&-1.3546\\2.3531&-0.8120\\0.9047&-0.3057
		\end{bmatrix}
		\\
		\widetilde{\Phi}_3&=
		\begin{bmatrix}			0.4667&0.3628\\-0.7651&-0.7875\\-0.6597&-1.3277\\1.5073&-0.6405
		\end{bmatrix}\!,\hspace{0.25cm}
\widetilde{\Phi}_4 =
		\begin{bmatrix}
-0.4462 & 0.3897 \\ -0.9077 & -0.3658 \\ 0.1943 & 0.9004 \\ 0.6469 & -0.9311
		\end{bmatrix}
	\end{align*}
\end{small}

With each random SCM sample $\widetilde{\Phi}_{k}$, an SCM series  $\widetilde{\Phi}_{k}(\tau)$ is generated according to the following formula,
\begin{equation}\label{Phi_tau}
\widetilde{\Phi}_{k}(\tau)=\Phi_0+\tau\cdot(\widetilde{\Phi}_{k}-\Phi_0)
\end{equation}
in which $k=1,2,3,4$, and $\tau$ uniformly distributes over the interval $[0.0000,\; 20.0000]$ with a constant distance $0.1000$. These mean that for each $k\in\{1,2,3,4\}$, 201 SCM samples are produced. In order to make the simulations meaningful, among these generated SCMs, an SCM that leads to an unstable NDS or does not satisfy the regularity assumption, is gotten rid of.

To measure the distance between the NDS ${\rm\bf\Sigma}(\Phi_0)$ and the NDS ${\rm\bf\Sigma}(\widetilde{\Phi}_k(\tau))$, let $\varepsilon_{k,\tau}(t)$ denote their external output difference under the same PRBS stimulations. That is, $\varepsilon_{k,\tau}(t)=y(t,\widetilde{\Phi}_{k}(\tau))-y(t,\Phi_0)$. Moreover, define a time domain distance $d^{[T]}_{k,\tau}$ and a frequency domain distance $d^{[F]}_{k,\tau}$ respectively as follows,
\begin{equation*}
	\begin{aligned}
		& d^{[T]}_{k,\tau}=\frac{1}{M}\sum^{M-1}_{t=0}\sqrt{\varepsilon_{k,\tau}^T(t)\varepsilon_{k,\tau}(t)}\\
		&
d^{[F]}_{k,\tau}= \parallel H(\lambda,\widetilde{\Phi}_{k}(\tau))-H(\lambda,\Phi_0)
		\parallel_\infty
	\end{aligned}
\end{equation*}
in which $M$ denotes the time domain sample number, while  $\parallel\cdot\parallel_\infty$ the $\mathcal{H}_\infty$ norm of a TFM. In addition, a quantity $d^{[S]}_{k,\tau}$ is also adopted, which is defined as
\begin{displaymath}
d^{[S]}_{k,\tau}=\inf_{\Phi\in\mathscr{U}(\Phi_0)}\overline{\sigma}(\widetilde{\Phi}_{k}(\tau)-\Phi)
\end{displaymath}
and directly measures the distance of an SCM $\widetilde{\Phi}_{k}(\tau)$ to the undifferentiable region $\mathscr{U}(\Phi_0)$ of the SCM $\Phi_0$.

From these definitions, it is clear that $d^{[T]}_{k,\tau}$ is the arithmetic mean for the Euclidean norm of the external output difference at each sampled time instant. Recall that the $\mathcal{H}_\infty$ norm of a TFM is defined as the supremum of the maximum singular value of its frequency responses and  extensively known as an induced norm \cite{zdg1996,zyl2018}, while the maximum singular value of a matrix is also an induced norm that has been widely used to measure matrix differences \cite{hj1991}. Each of these quantities appears to be a reasonable measure for NDS differences, and similar metrics have been adopted in \cite{Zhou2020-1}.

It may be understood without significant difficulties that for each $k\in \{1,2,3,4\}$, $d^{[S]}_{k,\tau}$ increases monotonically with $\tau$. As a matter of fact, it can even be rigorously proven for this particular example that the increment is linear. While these relations may not be trivial in NDS analysis and synthesis, the proof is omitted due to space considerations. On the other hand, these relations imply that variations of the NDS difference measured by $d^{[T]}_{k,\tau}$ or $d^{[F]}_{k,\tau}$ with respect to the SCM distance $d^{[S]}_{k,\tau}$ are proportional to its variations with the parameter $\tau$. Hence, only the latter are studied in the rest of this section.

\begin{figure}[!htb]
	\centering
		\begin{minipage}{0.48\columnwidth}
			\includegraphics[width=\linewidth]{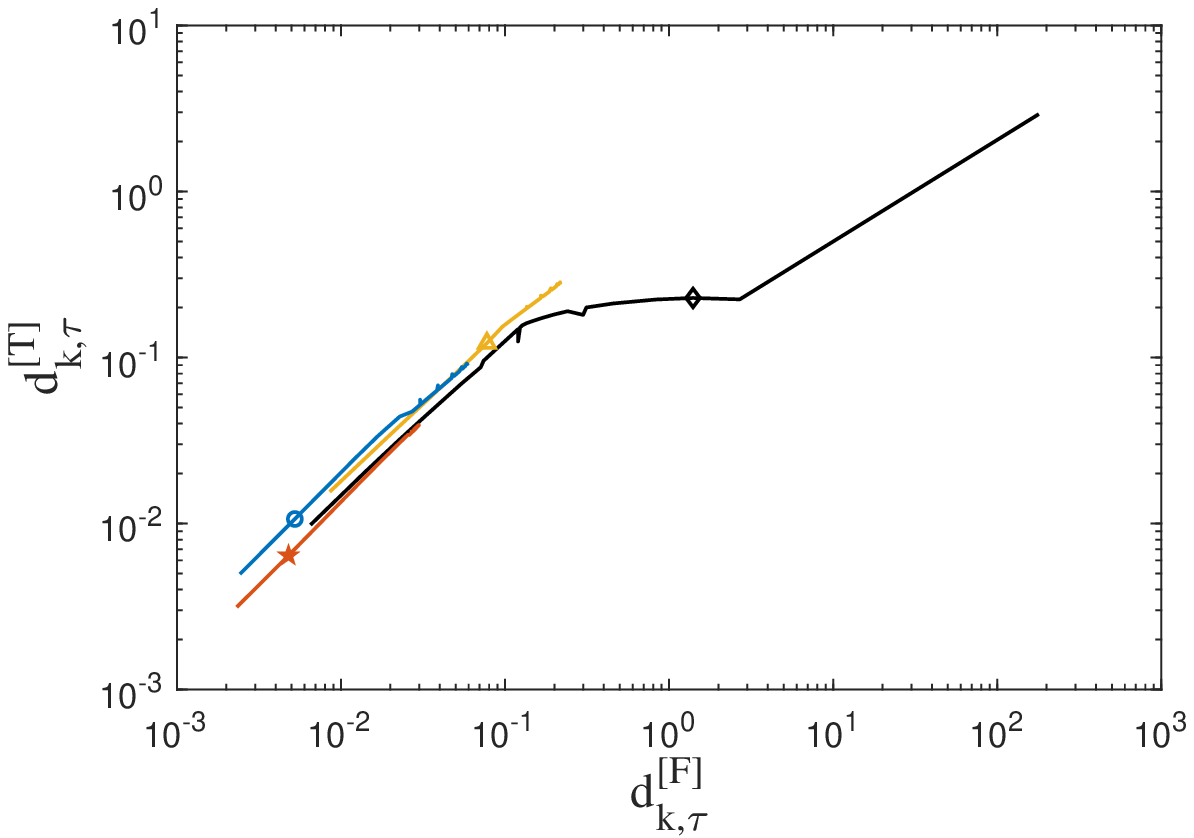}
{\small \hspace*{1.0cm} (a) $d^{[T]}_{k,\tau}$ {\it vs} $d^{[F]}_{k,\tau}$. }
	\end{minipage}
		\begin{minipage}{0.48\columnwidth}
			\includegraphics[width=\linewidth]{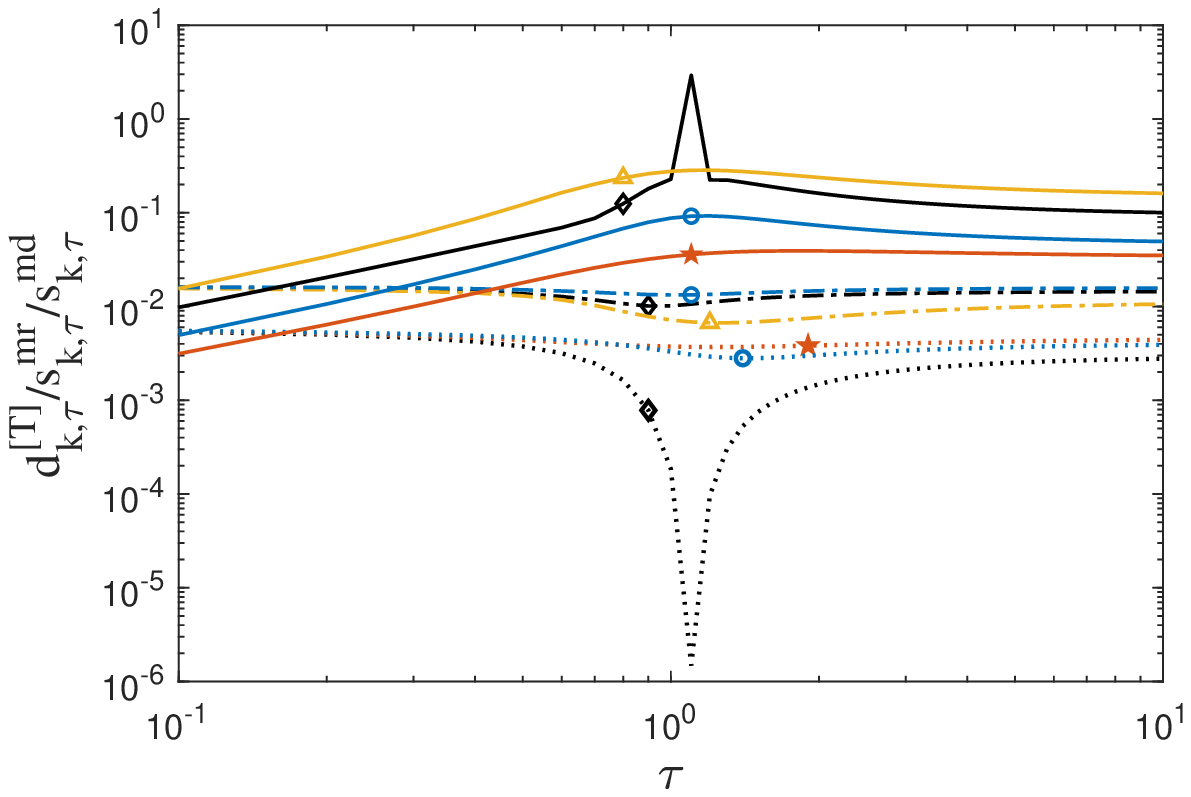}
{\small (b) stability margin and $d^{[T]}_{k,\tau}$. $-\!-$: $d^{[T]}_{k,\tau}$; $-\cdot-$: $s_{k,\tau}^{mr}$; $\cdots$: $s_{k,\tau}^{md}$.}
	\end{minipage}
	\caption{NDS differences measured in the time and frequency domains. $\Diamond$: $\widetilde{\Phi}_1$; $\triangle$: $\widetilde{\Phi}_2$; $\star$: $\widetilde{\Phi}_3$; $\circ$: $\widetilde{\Phi}_4$.}
	\label{figure2-1}
\end{figure}

Computation results are shown in Fig.$\ref{figure2-1}$. Fig.\ref{figure2-1}a shows variations of $d^{[T]}_{k,\tau}$ with respect to $d^{[F]}_{k,\tau}$. Obviously, for each $k=1,2,3,4$, $d^{[T]}_{k,\tau}$ increases almost monotonically with $d^{[F]}_{k,\tau}$, but the increment speed decreases very fast when $d^{[F]}_{k,\tau}$ takes a large value, recalling that the probing signals are two independent PRBSs which are in principle a random process. These observations may imply that the adopted time domain and the frequency domain measures on NDS distances, that is, $d^{[T]}_{k,\tau}$ and  $d^{[F]}_{k,\tau}$, are consistent well with each other, and a PRBS is in general good for NDS structure identification. These phenomena have also been observed in \cite{Zhou2020-1}.

Variations of $d^{[T]}_{k,\tau}$ with respect to the parameter $\tau$ are shown in Fig.\ref{figure2-1}b. Obviously, $d^{[T]}_{k,\tau}$ increases monotonically when $\tau$ is small, but decreases after $\tau$ reaches a certain value. These observations may imply that when an SCM $\Phi$ is close to the undifferentiable region $\mathscr{U}(\Phi_0)$, differences in the external outputs between the NDSs ${\rm\bf\Sigma}(\Phi)$ and ${\rm\bf\Sigma}(\Phi_{0})$ increase with its distance to the set $\mathscr{U}(\Phi_0)$, and this is not true when an SCM $\Phi$ is far away from $\mathscr{U}(\Phi_0)$.

Our analyses show that the peaks in the curves of $d^{[T]}_{k,\tau}$ of Fig.\ref{figure2-1}b may have a close relation with the NDS stability margin. To demonstrate these, let $s^{md}_{k,\tau}$ denote the minimum damping ratio of the complex eigenvalues of the STM of the NDS ${\rm\bf\Sigma}(\widetilde{\Phi}_{k}(\tau))$, while $s^{mr}_{k,\tau}$ the minimum absolute value of its real eigenvalues. These two quantities are adopted to measure the stability margin of an NDS whose engineering significance is clear from system and control theories, noting that a stable complex eigenvalue with a low damping ratio usually results in a long settling time and severe oscillations, while a stable real eigenvalue with a small absolute value corresponds to a large time constant and leads to a slow response \cite{Kailath1980,zdg1996}.

To reveal relations between an NDS stability margin and NDS differences,  Fig.\ref{figure2-1}b also draws scaled values of the aforementioned two stability margins. The scales are introduced to improve clarity, with which $s^{mr}_{k,\tau}$ and $s^{md}_{k,\tau}$ are multiplied respectively by $0.0160$ and $0.0020$. As the STM of the NDS  $\Sigma(\widetilde{\Phi}_2(\tau))$ has only complex eigenvalues for all the sampled $\tau$s, while the value of $s^{md}_{k,\tau}$ of the NDS  $\Sigma(\widetilde{\Phi}_3(\tau))$ is almost equal to zero for each $\tau$ sample, the related results are not very helpful in illustrating the influences of a stability margin on NDS differences. The associated curves are therefore not drawn in this figure and the following Fig.\ref{figure3-1}a.

A comparison of the positions of the peaks in Fig.\ref{figure2-1}b reveals that they are associated approximately with the same values of the parameter $\tau$. To show more clearly the influences of the NDS stability margin on the external output differences, variations of the time domain metric $d^{[T]}_{k,\tau}$ is plotted in Fig.\ref{figure3-1}a with respect to the aforementioned two stability margins, in which $d^{[T]}_{2,\tau}$ and $d^{[T]}_{4,\tau}$ are multiplied respectively by $0.1000$ and $10.0000$, once again in order to improve clarity. Obviously, with the increment of the stability margin, the time domain NDS distance really decreases almost monotonically.

\begin{figure}[!htb]
	\centering
		\begin{minipage}{0.48\columnwidth}
			\includegraphics[width=\linewidth]{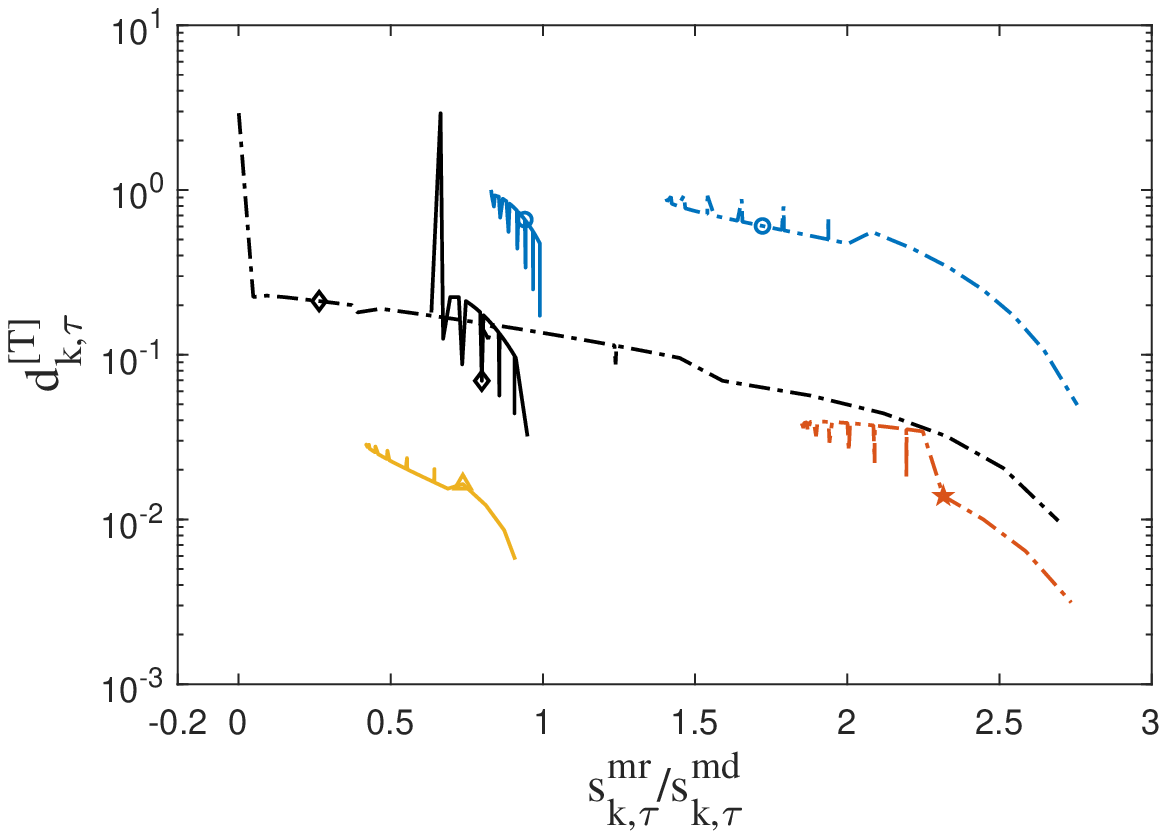}
{\small (a) $d^{[T]}_{k,\tau}$ {\it vs} the stability margins. $-\cdot-$: $s_{k,\tau}^{mr}$; $-\!-$: $s_{k,\tau}^{md}$.}
	\end{minipage}
		\begin{minipage}{0.48\columnwidth}
			\includegraphics[width=\linewidth]{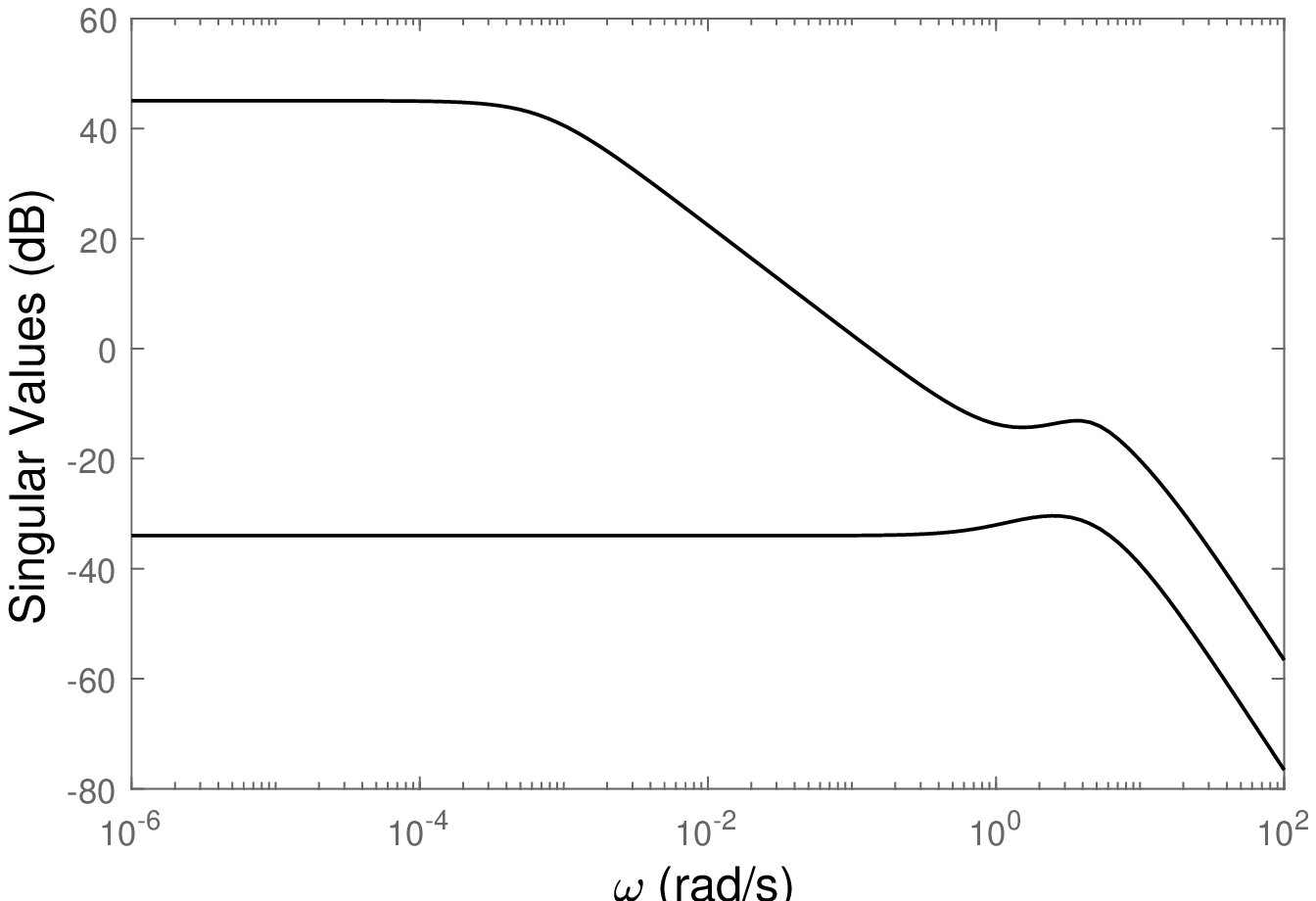}
{\small \hspace*{0.3cm} (b) NDS singular values.}
	\end{minipage}
	\caption{Relations among NDS differences, stability margin and NDS directions. $\Diamond$: $\widetilde{\Phi}_1$; $\triangle$: $\widetilde{\Phi}_2$; $\star$: $\widetilde{\Phi}_3$; $\circ$: $\widetilde{\Phi}_4$.}
	\label{figure3-1}
\end{figure}

An interesting phenomenon in Figs.\ref{figure2-1}b and \ref{figure3-1}a is that there are some zigzags in their curves, which means that when the parameter $\tau$ or the stability margin increases monotonically and continuously, there may exist some abrupt and significant changes in NDS external output differences. Primary analyses show that this phenomenon is closely related to directionality of a multi-input multi-output system which is well known in systems and control theories \cite{zdg1996,zyl2018}, but its effects appear to have not been extensively studied in system identifications \cite{adm2020,Ljung1999}.

To be more specific, $d^{[T]}_{1,\tau}$ is approximately equal to $1.0000\times 10^{-4}$ at which its curve in Fig.\ref{figure2-1}b takes a sharp decrement. At this value, the corresponding $d^{[F]}_{1,\tau}$ takes its maximum $1.7920\times10^2$, which is far greater than the $\mathcal{H}_\infty$ norm of the TFM $H(\lambda,\widetilde{\Phi}_{1}(\tau))-H(\lambda,\Phi_0)$ at other $\tau$s. Fig.\ref{figure3-1}b shows the singular values of the frequency response of the associated $H(\lambda,\widetilde{\Phi}_{[1]}(\tau))-H(\lambda,\Phi_0)$. Obviously, its condition number is quite large in the whole low frequency band, meaning that the associated external output differences are quite sensitive to the direction of an external input signal. These can also explain the zigzag phenomena of Fig.\ref{figure3-1}a. These observations may imply that independent PRBSs are not always very efficient in NDS structure identification.

\section{Concluding Remarks}

This paper investigates requirements on an NDS such that its SCM can be estimated from experiment data, under the condition that its subsystem dynamics are described in a descriptor form. Except the assumption that the NDS itself and its subsystems are regular, which is essential for the NDS to work properly, not any other assumptions are adopted. An algebraic criterion is established for the global structure identifiability of the NDS with some specific subsystem connections, which is in principle computationally feasible, at least for an NDS with a moderate number of subsystems. Undifferentiable regions have also been explicitly described for a particular set of unidentifiable subsystem connections, which reveals that global and local identifiability are equivalent to each other for the structure of this NDS. Several particular situations have also been discussed in which some a priori information is available for an NDS structure, and similar results have been obtained. Subsystem interaction reconstruction has also been studied under the condition that the NDS is well-posed. Necessary and sufficient conditions have been established respectively for a time domain model being consistent with the NDS structure, and the NDS structure is reconstructible from a time domain model.

Numerical simulation results indicate that when an SCM is close to an undifferentiable region, NDS external output differences increase monotonically with the deviation of the SCM to the undifferentiable region. However, when an SCM is close to the set of subsystem connections that lead to an unstable NDS, rather than the distance between the associated SCMs, NDS external output differences appear to be dominated by the NDS stability margin. It has also been observed from these numerical studies that independent PRBSs may not always be very efficient in probing NDS structures.

As a further issue, it is interesting to develop a condition for NDS structure identifiability that has a lower computational cost for a large scale NDS with algebraically dependent subsystem interactions. In addition, probing signal designs are also worthwhile to study with some more efforts.

\renewcommand{\theequation}{A\arabic{equation}}
\setcounter{equation}{0}
\small
\section*{Appendix: Proof of Some Technical Results}

% \vspace{0.25cm}
\noindent\textbf{Proof of Theorem $\ref{theo1}$:} Assume that there is a $\Phi_\star \in \mathscr{S}$ which is different from $\Phi_0$ and satisfies $H(\lambda,\Phi_\star)-H(\lambda,\Phi_0) \equiv 0$. Then according to the definition of the TFM $H(\lambda,\Phi)$ given in Eq.(\ref{tfm}), we have that
\begin{eqnarray}
& & H(\lambda,\Phi_\star)-H(\lambda,\Phi_0) \nonumber \\
&=& G_{yv}(\lambda)\left\{\Phi_\star[I_{m_z}-G_{zv}(\lambda)\Phi_\star]^{-1}- \right.\nonumber \\
& & \hspace*{2.5cm}\left.[I_{m_v}-\Phi_0G_{zv}(\lambda)]^{-1}\Phi_0\right\}G_{zu}(\lambda)\nonumber \\
&=& G_{yv}(\lambda)[I_{m_v}-\Phi_0G_{zv}(\lambda)]^{-1}\Delta_\star \times \nonumber\\
& & \hspace*{2.5cm} [I_{m_z}-G_{zv}(\lambda)\Phi_\star]^{-1}G_{zu}(\lambda) \nonumber \\
&\equiv&  0
\label{a.11}
\end{eqnarray}
in which $\Delta_\star=\Phi_0-\Phi_\star$. According to Lemma $\ref{lemma2}$, the structure of the NDS $\rm\bf\Sigma$ is not globally identifiable at this specific SCM $\Phi_0$, if and only if there is a real matrix $\Delta_\star\not=0$ satisfying the above equation.

For each $i=1,2,\cdots,N$, denote by $r_{zu}^{[i]}$ the normal rank of the TFM $G_{zu}(\lambda,i)$. Write its Smith-McMillan form as
\begin{equation*}
	G_{zu}(\lambda,i)=U_{zu}(\lambda,i)
	\left[\begin{array}{cc}
diag\{\kappa^{[j]}_{zu}(\lambda,i)|^{r^{[i]}_{zu}}_{j=1}\} & 0 \\ 0 & 0
	\end{array}     \right]
	V^{T}_{zu}(\lambda,i)
\end{equation*}
For each $\star=z, y$, $\#=v, u$ and $i=1,2,\cdots,N$, partition the unimodular MVP $U_{\star\#}(\lambda,i)$ into two column blocks as
$U_{\star\#}(\lambda,i) = \left[
		U^{[1]}_{\star\#}(\lambda,i)\;\;
        U^{[2]}_{\star\#}(\lambda,i) \right]$,
in which the sub-MVP $U^{[1]}_{\star\#}(\lambda,i)$ has  $r_{\star\#}^{[i]}$ columns. Moreover, denote
the inverse matrix of this unimodular MVP by  $U^{[iv]}_{\star\#}(\lambda,i)$, and partition it into two row blocks. That is, $U_{\star\#}^{-1}(\lambda,i) = U^{[iv]}_{\star\#}(\lambda,i)=
	col\left\{
		U^{[iv,1]}_{\star\#}(\lambda,i),\;
        U^{[iv,2]}_{\star\#}(\lambda,i) \right\}$,
in which $U^{[iv,1]}_{\star\#}(\lambda,i)$ has $r_{\star\#}^{[i]}$ rows. Furthermore, define an integer $r_{\star\#}$ and two MVPs $U^{[k]}_{\star\#}(\lambda)$ and $U^{[iv,k]}_{\star\#}(\lambda)$ with $k\in \{1,2\}$ respectively as   $r_{\star\#}=\sum^N_{i=1}r^{[i]}_{\star\#}$, $U^{[k]}_{\star\#}(\lambda)=diag\{U^{[k]}_{\star\#}(\lambda,i)|^N_{i=1}\}$
and  $U^{[iv,k]}_{\star\#}(\lambda)=diag\{U^{[iv,k]}_{\star\#}(\lambda,i)|^N_{i=1}\}$.
Then from the consistent block diagonal structures of the involved MVPs, straightforward algebraic manipulations show that $U_{\star\#}^{[iv,1]}(\lambda)U_{}^{[1]}(\lambda) \!\equiv\!  I_{r_{\star\#}}$ and
\begin{equation}
U_{\star\#}^{[iv,2]}(\lambda)U_{\star\#}^{[1]}(\lambda) \!\equiv\! 0_{(m_{\star} \!-\! r_{\star\#})\times r_{\star\#}}
\label{a.13.1}
\end{equation}

Recall that the TFM $G_{zu}(\lambda)$ is block diagonal. It can be directly shown from the above relations that  $U^{[iv,2]}_{zu}(\lambda)G_{zu}(\lambda) \equiv 0$. Therefore, the satisfaction of Eq.(\ref{a.11}) is equivalent to the existence of a $m_y\times(m_z-r_{zu})$ dimensional RFM  $\Pi_1(\lambda)$ such that the following equality is satisfied,
\begin{align}\label{a.12} &G_{yv}(\lambda)[I_{m_v}-\Phi_0G_{zv}(\lambda)]^{-1}\Delta_\star[I_{m_z}-G_{zv}(\lambda)\Phi_\star]^{-1}\notag\\
	=&\Pi_1(\lambda)U^{[iv,2]}_{zu}(\lambda)
\end{align}

Note that for each $i=1,2,\cdots,N$, the MVP $V_{yv}(\lambda,i)$ of Eq.(\ref{SM}) is also unimodular. We therefore have that its inverse $V_{yv}^{[iv]}(\lambda,i)$ is also an unimodular MVP.
Partition the unimodular MVPs $V_{yv}(\lambda,i)$ and $V^{[iv]}_{yv}(\lambda,i)$ respectively as $V_{yv}(\lambda,i)=\left[V^{[1]}_{yv}(\lambda,i)\quad V^{[2]}_{yv}(\lambda,i)\right]$ and $V^{[iv]}_{yv}(\lambda,i)=col\left\{V^{[iv,1]}_{yv}(\lambda,i),\quad V^{[iv,2]}_{yv}(\lambda,i)\right\}$, in which $V^{[1]}_{yv}(\lambda,i)$ has $r_{yv}^{[i]}$ columns and $V^{[iv,1]}_{yv}(\lambda,i)$ has $r_{yv}^{[i]}$ rows. Moreover, for $k=1$ and $2$, define MVPs $V^{[k]}_{yv}(\lambda)$ and $V^{[iv,k]}_{yv}(\lambda)$, as well as a RFM $\Lambda_{yv}(\lambda)$, respectively as  $V^{[k]}_{yv}(\lambda) = diag\{V^{[k]}_{yv}(\lambda,i)|^N_{i=1}\}$, $V^{[iv,k]}_{yv}(\lambda) = diag\{V^{[iv,k]}_{yv}(\lambda,i)|^N_{i=1}\}$, and  $\Lambda_{yv}(\lambda) = diag\{[diag\{\kappa^{[j]}_{yv}(\lambda,i)|^{r^{[i]}_{yv}}_{j=1}\}]|^N_{i=1}\}$. From these definitions, it is obvious that $V_{yv}^{[1]T}(\lambda)V_{yv}^{[iv,1]T}(\lambda)\equiv I_{r_{yv}}$ and the following equality holds,
\begin{equation}
V_{yv}^{[1]T}(\lambda)V_{yv}^{[iv,2]T}(\lambda)\equiv 0_{(m_{v} \!-\! r_{yv})\times r_{yv}}
\label{a.13.2}
\end{equation}

On the basis of these relations and the Smith-McMillan form of the TFM $G_{yv}(\lambda,i)$, Eq.(\ref{a.12}) can be equivalently rewritten as
the following two equations,
\begin{eqnarray}
& & \Lambda_{yv}(\lambda)V_{yv}^{[1]T}(\lambda)[I_{m_v}-\Phi_0G_{zv}(\lambda)]^{-1}\Delta_\star=  \nonumber\\ & & \hspace*{1.5cm} U^{[iv,1]}_{yv}(\lambda)\Pi_1(\lambda)U^{[iv,2]}_{zu}(\lambda)[I_{m_z}-G_{zv}(\lambda)\Phi_\star] \label{a.19} \\
& & U^{[iv,2]}_{yv}(\lambda)\Pi_1(\lambda)U^{[iv,2]}_{zu}(\lambda)[I_{m_z}-G_{zv}(\lambda)\Phi_\star]\equiv0
\label{a.20}
\end{eqnarray}

From Eq.(\ref{a.13.1}), it is obvious that the satisfaction of Eq.(\ref{a.20}) is equivalent to the existence of a $r_{yv}\times m_z$ dimensional RFM $\Pi_2(\lambda)$, such that
\begin{displaymath}
\Pi_1(\lambda)U^{[iv,2]}_{zu}(\lambda)[I_{m_z}-G_{zv}(\lambda)\Phi_\star]=U_{yv}^{[1]}(\lambda)\Pi_2(\lambda)
\end{displaymath}
Substitute this relation back into Eq.(\ref{a.19}), the following equality is obtained,
\begin{equation}
V_{yv}^{[1]T}(\lambda)[I_{m_v}-\Phi_0G_{zv}(\lambda)]^{-1}\Delta_\star=\Lambda^{-1}_{yv}(\lambda)\Pi_2(\lambda)
\end{equation}

From this equality and Eq.(\ref{a.13.2}), it can be claimed that
there is a $(m_v-r_{yv})\times m_z$ dimensional RFM $\Pi_3(\lambda)$ satisfying the following equation,
\begin{eqnarray}\label{a.15} & & [I_{m_v}-\Phi_0G_{zv}(\lambda)]^{-1}\Delta_\star \nonumber \\
&=& V_{yv}^{[iv,1]T}(\lambda)\overline{\Pi}_2(\lambda) + V_{yv}^{[iv,2]T}(\lambda)\Pi_3(\lambda)
\end{eqnarray}
in which $\overline{\Pi}_2(\lambda)=\Lambda^{-1}_{yv}(\lambda)\Pi_2(\lambda)$ that defines a bijective mapping between the RFM $\overline{\Pi}_2(\lambda)$ and the RFM $\Pi_2(\lambda)$.

Denote the RFM $col\{\overline{\Pi}_2(\lambda),\; \Pi_3(\lambda)\}$ by $\Pi(\lambda)$. Then the aforementioned equation can be rewritten as
\begin{equation*}
[I_{m_v}-\Phi_0G_{zv}(\lambda)]^{-1}\Delta_\star = V^{[iv]T}_{yv}(\lambda)\Pi(\lambda)
\end{equation*}
Denote the block diagonal MVPs  $diag\{R(\lambda,i)|^N_{i=1}\}$, $diag\{Q(\lambda,i)|^N_{i=1}\}$, $diag\{\Omega(\lambda,i)|^N_{i=1}\}$, $diag\{N_{zv}(\lambda,i)|^N_{i=1}\}$ and  $diag\{D_{zv}(\lambda,i)|^N_{i=1}\}$ respectively by $R(\lambda)$, $Q(\lambda)$,  $\Omega(\lambda)$, $N_{zv}(\lambda)$ and $D_{zv}(\lambda)$. Then Eq.(\ref{a.15}) is equivalent to
\begin{align}\label{a.16} \Delta_\star&=[I_{m_v}-\Phi_0G_{zv}(\lambda)]V_{yv}^{[iv]}(\lambda)\Pi(\lambda)\notag\\
&=[D_{zv}(\lambda)-\Phi_0N_{zv}(\lambda)]D^{-1}_{zv}(\lambda)V_{yv}^{[iv]T}(\lambda)\Pi(\lambda)\notag\\
&=[D_{zv}(\lambda)-\Phi_0N_{zv}(\lambda)][R(\lambda)+D^{-1}_{zv}(\lambda)\overline{V}_{yv}^{[iv]T}(\lambda)]\Pi(\lambda)\notag\\
&=[D_{zv}(\lambda)-\Phi_0N_{zv}(\lambda)][R(\lambda)+Q(\lambda)\Omega^{-1}(\lambda)]\Pi(\lambda)\notag\\
	&=[X(\lambda)-\Phi_0Y(\lambda)]\overline{\Pi}(\lambda)
\end{align}
in which $\overline{\Pi}(\lambda)=\Omega^{-1}(\lambda)\Pi(\lambda)$. This RFM can be assumed without any loss of generality to be a MVP that possibly has an infinite degree. Once again, the mapping between the RFM $\overline{\Pi}(\lambda)$ and the RFM $\Pi(\lambda)$ is bijective. Besides, $D^{-1}_{zv}(\lambda)\overline{V}_{yv}^{[iv]}(\lambda)$ and $Q(\lambda)\Omega^{-1}(\lambda)$ are respectively a left MFD and a right MFD of the strict RFM $D^{-1}_{zv}(\lambda)V_{yv}^{[iv]T}(\lambda) - R(\lambda)$.

For each $j=1,\cdots,m_z$, let $\mathbf{\delta}_{\star,j}$ and $\mathbf{\pi}_{j}(\lambda)$ denote respectively the $j$-th column vector of the matrix $\Delta_\star$ and the RFM  $\overline{\Pi}(\lambda)$. Obviously, $\mathbf{\delta}_{\star,j} \in \mathcal{R}^{m_v}$ and  $\mathbf{\pi}_{j}(\lambda)$ is an ${m_v}$ dimensional VVP that may have an infinite degree. Note that the existence of a nonzero real matrix $\Delta_\star$ satisfying Eq.(\ref{a.16}) is equivalent to the existence of a nonzero real column vector $\mathbf{\delta}_{\star,j}$, such that the equation $[X(\lambda)-\Phi_0 Y(\lambda)]\mathbf{\alpha}_{j}(\lambda) = \mathbf{\delta}_{\star,j}$ has a solution.

This proves the converse negative proposition of this theorem, and therefore completes the proof.   \hspace{\fill}$\Diamond$

\vspace{0.5cm}
\noindent\textbf{Proof of Theorem $\ref{theo2}$:} Partition the unimodular MVP $V_{\Phi_0}(\lambda)$ of Eq.(\ref{Smith}) into $\left[V^{[1]}_{\Phi_0}(\lambda) \quad V^{[2]}_{\Phi_0}(\lambda)\right]$, in which $V^{[1]}_{\Phi_0}(\lambda)$ has $r(\Phi_0)$ columns. Moreover, define a MVF $\Lambda_{\Phi_0}(\lambda)$ as  $\Lambda_{\Phi_0}(\lambda)=diag\{\mu^{[j]}_{\Phi_0}(\lambda)|^{r(\Phi_0)}_{j=1}\}$. Then Eq.(\ref{Smith}) can be equivalently rewritten as follow,
\begin{equation}\label{a.17}
	\begin{bmatrix}	\Lambda_{\Phi_0}(\lambda)V^{[1]T}_{\Phi_0}(\lambda)\alpha(\lambda)\\0
	\end{bmatrix}
	=
	\begin{bmatrix}
		U^{[iv,1]}_{\Phi_0}(\lambda)\\   U^{[iv,2]}_{\Phi_0}(\lambda)
	\end{bmatrix}
	\delta
\end{equation}
which are actually equal to the following two equations,
\begin{eqnarray}
& & \Lambda_{\Phi_0}(\lambda)V^{[1]}_{\Phi_0}(\lambda)\alpha(\lambda) = U^{[iv,1]}_{\Phi_0}(\lambda)\delta
\label{a.21} \\
& & U^{[iv,2]}_{\Phi_0}(\lambda)\delta \equiv 0
\label{a.22}
\end{eqnarray}

Denote the inverse of the unimodular MVP $V_{\Phi_0}(\lambda)$ by  $V^{[iv]}_{\Phi_0}(\lambda)$, and partition it as  $V^{[iv]}_{\Phi_0}(\lambda) = col\left\{V^{[iv,1]}_{\Phi_0}(\lambda),  \right.$ $\left. V^{[iv,2]}_{\Phi_0}(\lambda)\right\}$, in which $V^{[iv,1]}_{\Phi_0}(\lambda)$ has $r(\Phi_0)$ columns. From their definitions, we immediately have the equalities $V^{[1]T}_{\Phi_0}V^{[iv,1]T}_{\Phi_0} \equiv I_{r(\Phi_0)}$ and  $V^{[1]T}_{\Phi_0}V^{[iv,2]}_{\Phi_0} \equiv 0_{r(\Phi_0) \times (m_{v}-r(\Phi_0))}$. Therefore, for an arbitrary vector $\delta$ with a compatible dimension,
\begin{displaymath}
\alpha(\lambda)= V_{\Phi_0}^{[iv,1]T}(\lambda)\Lambda^{-1}_{\Phi_0}(\lambda)U_{\Phi_0}^{[iv,1]}\delta + V_{\Phi_0}^{[iv,2]T}(\lambda)\pi(\lambda)
\end{displaymath}
satisfies Eq.(\ref{a.21}), in which $\pi(\lambda)$ is an arbitrary $m_v-r(\Phi_0)$ dimensional VVP that may have an infinite degrees. On the contrary, it can also be straightforwardly shown that any VVP $\alpha(\lambda)$ satisfying Eq.(\ref{a.21}) can be expressed by this formula. These arguments mean that there is a solution to Eq.(\ref{a.17}), if and only if Eq.(\ref{a.22}) has a solution.

On the other hand, Eq.(\ref{a.22}) means that the real vector $\delta$ must be a constant element in the right null space of the MVP  $U_{\Phi_0}^{[iv,2]}(\lambda)$ which has a finite degree. This equation can be equivalently rewritten as follow,
\begin{eqnarray}	
U^{[iv,2]}_{\Phi_0}(\lambda)\delta &=& \left(\sum^{p(\Phi_0)}_{k=0}\lambda^k U_{\Phi_0,k}^{[iv,2]}\right)\delta  \nonumber \\
&=& \left[1\quad\lambda\cdots\lambda^p\right])U_{\Phi_0,k}^{[iv,2]} \delta \nonumber\\
&\equiv& 0    \label{a.18}
\end{eqnarray}
which is further equivalent to
\begin{equation}
U_{\Phi_0,k}^{[iv,2]}\delta = 0
\label{a.23}
\end{equation}

According to Theorem $\ref{theo1}$, the structure of the NDS $\rm\bf\Sigma$ is not globally identifiable at this specific SCM $\Phi_{0}$, if and only if there is a nonzero vector $\delta\in\mathcal{R}^{m_v}$ satisfies Eq.(\ref{lemma}), and thus satisfies Eq.(\ref{a.23}). This proves the converse negative proposition of this theorem, and thus completes the proof.   \hspace{\fill}$\Diamond$

\vspace{0.5cm}
\noindent\textbf{Proof of Theorem $\ref{theo4}$:} For brevity, define  matrices $H$, $G$, $K$ and $L$ respectively as
\small{
\begin{equation*}
H \!=\! \begin{bmatrix}
		\widehat{A} & \widehat{B} \\ \widehat{C} & \widehat{D}
	\end{bmatrix}\!\!, \hspace{0.2cm}
	G \!=\!
	\begin{bmatrix}
	A_{xx} & B_{xu} \\
	C_{yx} & D_{yu}
	\end{bmatrix}\!\!, \hspace{0.2cm}
    K \!=\!
	\begin{bmatrix}
		B_{xv} \\ D_{yv}
	\end{bmatrix}\!\!, \hspace{0.2cm}
	L \!=\!
	\begin{bmatrix}
		C_{zx} & D_{zu}
	\end{bmatrix}
\end{equation*}}

When a descriptor form model ${\rm\bf\Sigma}(\widehat{A}:\widehat{E})$ is available for the whole NDS $\rm\bf\Sigma$ that is consistent with the NDS structure, on the basis of Eq.(\ref{SSM}), we have that there is at least one SCM $\Phi$ such that the following equality is satisfied,
\begin{equation}\label{SSM2}
	H=G+K(I_{m_v}-\Phi D_{zv})^{-1}\Phi L
\end{equation}

Assume that the structure of the NDS $\rm\bf\Sigma$ is reconstructible from its descriptor form model. Moreover, assume that the ranks of the matrices $K$ and $L$ are equal to $r_K$ and $r_L$  respectively. Then their singular value decompositions can in general be expressed as follows,
\begin{align}
	K=&
	\begin{bmatrix}
		U_K^{[1]} & U_K^{[2]}
	\end{bmatrix}
	\begin{bmatrix}
		\Lambda_K & 0 \\ 0 & 0
	\end{bmatrix}
    \begin{bmatrix}
    	V_K^{[1]} & V_K^{[2]}
    \end{bmatrix}^T  \label{a.24}  \\
	L=&
	\begin{bmatrix}
		U_L^{[1]} & U_L^{[2]}
	\end{bmatrix}
	\begin{bmatrix}
		\Lambda_L & 0 \\ 0 & 0
	\end{bmatrix}
	\begin{bmatrix}
		V_L^{[1]} & V_L^{[2]}
	\end{bmatrix}^T \label{a.25}
\end{align}
in which for each $\star = L$ or $K$, $\Lambda_\star\in\mathcal{R}^{r_\star\times r_\star}$ is a diagonal matrix with positive diagonal elements, while $U^{[1]}_\star$ and  $V^{[1]}_\star$ have $r_\star$ columns, and $[U^{[1]}_\star\quad U^{[2]}_\star]$, $[V^{[1]}_\star\quad V^{[2]}_\star]$ are orthogonal  matrices with compatible dimensions. Denote $m_{x}+m_{y}$ and $m_{x}+m_{u}$ respectively by $m_{K}$ and $m_{L}$. From these decompositions, it is obvious that the following equalities hold for every $\star \in \{L,\;K \}$.
\begin{eqnarray}
& & U_\star^{[1]T}U_\star^{[1]} = I_{r_\star},\hspace{0.25cm} U_\star^{[2]T}U_\star^{[1]}=0_{(m_{\star}-{r_\star})\times r_\star}
\label{a.31}\\
& & V_\star^{[1]T}V_\star^{[1]} = I_{r_\star}, \hspace{0.25cm} V_\star^{[1]T}V_\star^{[2]}=0_{(m_{\star}-{r_\star})\times r_\star}
\label{a.32}
\end{eqnarray}

Define a matrix $H_{0}$ as  $H_0=\Lambda_K^{-1}U_K^{[1]T}(H-G)V_L^{[1]}\Lambda_L^{-1}$. Then it can be straightforwardly shown from Eq.(\ref{SSM2}) that
$H_0=V_K^{[1]T}(I_{m_v}-\Phi D_{zv})^{-1}\Phi U_L^{[1]}$. In addition, Eq.(\ref{SSM2}) can be equivalently rewritten as
\begin{equation}\label{a.3}
	(I_{m_v}-\Phi D_{zv})^{-1}\Phi=V_K^{[1]}(H_0U_L^{[1]T}+\Pi_1U_L^{[2]T})+V_K^{[2]}\Pi_2
\end{equation}
in which $\Pi_1$ and  $\Pi_2$ are arbitrary $r_K\times(m_z-r_L)$ and $(m_v-r_K)\times m_z$ dimensional real matrices.

Denote the matrix  $V_K^{[1]}(H_0U_L^{[1]^T}+\Pi_1U_L^{[2]T})+V_K^{[2]}\Pi_2$ by $\Pi$ for conciseness. Then Eq.(\ref{a.3}) can be rewritten as
\begin{equation}
\Phi \left[I_{m_v}+D_{zv}\Pi \right]=\Pi
\label{a.29}
\end{equation}
which has a unique solution for the SCM $\Phi$, if and only if the matrix $I_{m_v}+D_{zv}\Pi$ is invertible.

Assume now that the matrices $\Pi_1\in \mathcal{R}^{r_K\times(m_z-r_L)}$ and  $\Pi_2 \in \mathcal{R}^{(m_v-r_K)\times m_z}$ satisfy the requirement that the associated matrix $I_{m_v}+D_{zv}\Pi$ is invertible. This existence is always guaranteed, recalling that the descriptor form model ${\rm\bf\Sigma}(\widehat{A}:\widehat{E})$ of the whole NDS $\rm\bf\Sigma$ is assumed to be consistent with the NDS structure. More specifically, this consistency guarantees that there is at least one matrix pair $(\Pi_1,\;\Pi_2)$, such that there is a solution to Eq.(\ref{a.29}). If the corresponding matrix $I_{m_v}+D_{zv}\Pi$ is not invertible, then according to matrix theories \cite{hj1991}, there are infinitely many SCM $\Phi$s that satisfy Eq.(\ref{a.29}), and therefore Eq.(\ref{SSM2}). Hence, the structure of the NDS $\rm\bf\Sigma$ is not reconstructible from its descriptor form model, and is therefore a contradiction.

On the other hand, for arbitrary matrices $\Delta_{1}\in \mathcal{R}^{r_K\times(m_z-r_L)}$ and  $\Delta_{2}\in\mathcal{R}^{(m_v-r_K)\times m_z}$, define matrices $\overline{\Pi}_1$ and $\overline{\Pi}_2$ respectively as
$\overline{\Pi}_1 = \Pi_1 +\Delta_{1}$ and $\overline{\Pi}_2 = \Pi_2 +\Delta_{2}$. Moreover, define a matrix $\overline{\Pi}$ as
$\overline{\Pi} = V_K^{[1]}(H_0U_L^{[1]^T}+\overline{\Pi}_1 U_L^{[2]T})+V_K^{[2]}\overline{\Pi}_2$. Then
\begin{eqnarray}
I_{m_v}+D_{zv}\overline{\Pi} &=& \left(I_{m_v}+D_{zv}\Pi\right)\left\{I_{m_v}+ \left(I_{m_v}+ \right.\right.  \nonumber\\
& & \hspace*{0.25cm}\left. \left. D_{zv}\Pi\right)^{-1}
\left[V_K^{[1]}{\Delta}_1 U_L^{[2]T}+V_K^{[2]}{\Delta_2}
\right] \right\}
\label{a.30}
\end{eqnarray}

Based on matrix analysis \cite{hj1991}, direct algebraic manipulations show that $\rho_{max}\left\{\left(I_{m_v}+ D_{zv}\Pi\right)^{-1}
\left[V_K^{[1]}{\Delta}_1 U_L^{[2]T}+V_K^{[2]}{\Delta_2}
\right] \right\}$ $\leq
 \overline{\sigma}\left\{\left(I_{m_v}+ D_{zv}\Pi\right)^{-1}\right\}\{\overline{\sigma}(V_K^{[1]})
\overline{\sigma}(U_L^{[2]T})
\overline{\sigma}({\Delta}_1 ) + \overline{\sigma}( V_K^{[2]})\times $
$\overline{\sigma}({\Delta}_2 )\}$. It can therefore be declared that if the matrix $I_{m_v}+D_{zv}\Pi$ is invertible, then there exists a positive number $\varepsilon$, such that for each matrix pair $\left(\Delta_{1},\; \Delta_{2}\right)$ satisfying  $\Delta_{1}\in \mathcal{R}^{r_K\times(m_z-r_L)}$ with $\overline{\sigma}(\Delta_{1}) < \varepsilon$ and  $\Delta_{2}\in\mathcal{R}^{(m_v-r_K)\times m_z}$ with $\overline{\sigma}(\Delta_{2}) < \varepsilon$, the following inequality is satisfied,
\begin{equation}
\rho_{max}\left\{\left(I_{m_v}+ D_{zv}\Pi\right)^{-1}
\left[V_K^{[1]}{\Delta}_1 U_L^{[2]T}+V_K^{[2]}{\Delta_2}
\right] \right\} < 1
\end{equation}

This inequality and Eq.(\ref{a.30}) further mean that
the associated matrix $I_{m_v}+D_{zv}\overline{\Pi}$ is also invertible. Hence, Eq.(\ref{a.29}) also has a solution when the matrix ${\Pi}$ is replaced by the matrix $\overline{\Pi}$, whenever the aforementioned conditions are satisfied by the matrices $\Delta_{1}$ and $\Delta_{2}$.

Assume that there are two different matrix pairs $(\Pi_1,\Pi_2)$ and
$(\overline{\Pi}_1,\overline{\Pi}_2)$ simultaneously satisfying Eq.(\ref{a.3}). Then
\begin{eqnarray*}
& & V_K^{[1]}(\Pi_1-\overline{\Pi}_1)U_L^{[2]T}+V_K^{[2]}(\Pi_2-\overline{\Pi}_2) \\
&=& \left[V_K^{[1]}\;\; V_K^{[2]}\right]\left[\begin{array}{c}
(\Pi_1-\overline{\Pi}_1)U_L^{[2]T} \\  \Pi_2-\overline{\Pi}_2
\end{array}\right] \\
&=& 0
\end{eqnarray*}
Note that the matrix $\left[V_K^{[1]}\;\; V_K^{[2]}\right]$ is invertible, while the matrix $U_L^{[2]T}$ is of FRR. The last equality in the above equation means that $\overline{\Pi}_1=\Pi_1$ and  $\overline{\Pi}_2=\Pi_2$. Hence, the solution to Eq.(\ref{a.29}) is different with a distinctive matrix pair $(\Pi_1,\Pi_2)$, provided that a solution exists.

These arguments reveal that if the matrices $U_L^{[2]}$ and $V_K^{[2]}$ are not empty, then there are infinite SCM $\Phi$s that satisfy Eq.(\ref{SSM2}), meaning that in order to guarantee that the structure of the NDS $\rm\bf\Sigma$ is reconstructible from its descriptor form model ${\rm\bf\Sigma}(\widehat{A}:\widehat{E})$, it is necessary that the matrix $K$ is of FCR, while the matrix $L$ is of FRR.

Now, assume that the matrix $K$ is of FCR, and the matrix $L$ is of FRR.
Then both the matrices $K^{T}K$ and $LL^{T}$ are invertible. We therefore have the following relation from Eq.(\ref{SSM2}),
\begin{equation}\label{a.26}
(I_{m_v}-\Phi D_{zv})^{-1}\Phi=(K^{T}K)^{-1}K^{T}(H-G)L^{T}(LL^{T})^{-1}
\end{equation}

If there are two different SCM $\Phi$s satisfying the above equation, denote them respectively by $\Phi_{1}$ and $\Phi_{2}$. Then
\begin{eqnarray}\label{a.27}
(I_{m_v}-\Phi_{1} D_{zv})^{-1}\Phi_{1}
&=& (I_{m_v}-\Phi_{2} D_{zv})^{-1}\Phi_{2} \nonumber \\
&=& \Phi_{2}(I_{m_z} - D_{zv}\Phi_{2})^{-1}
\end{eqnarray}
which can be directly shown to be equivalent to
\begin{equation}\label{a.28}
(I_{m_v}-\Phi_{1} D_{zv})^{-1}(\Phi_{2}-\Phi_{1})(I_{m_z} - D_{zv}\Phi_{2})^{-1}=0
\end{equation}
Hence $\Phi_{2}=\Phi_{1}$. That is, there is at most one SCM $\Phi$ that satisfies Eq.(\ref{SSM2}), meaning that the structure of the NDS $\rm\bf\Sigma$ is  reconstructible from its descriptor form model.

This completes the proof.   \hspace{\fill}$\Diamond$

\vspace{0.5cm}
\noindent\textbf{Proof of Corollary $\ref{coro4}$:} In this proof, the matrices $H$ and $G$ are used again for brevity, which are defined in the proof of Theorem \ref{theo4}. From  Eq.(\ref{SSM}), we have that a descriptor form model ${\rm\bf\Sigma}(\widehat{A}:\widehat{E})$ is consistent with the structure of the NDS $\rm\bf\Sigma$, if and only if the following equation has at least one solution,
\begin{equation}\label{SSM2-x}
	E_{d} = K(I_{m_v}-\Phi D_{zv})^{-1}\Phi L
\end{equation}

On the other hand, when the structure of the NDS $\rm\bf\Sigma$ is reconstructible, we have from Theorem \ref{theo4} that the matrix $K$ is of FCR, while the matrix $L$ is of FRR. This means that both the matrices $K^{T}K$ and $LL^{T}$ are invertible.

Assume now that there is an SCM $\Phi$ satisfying Eq.(\ref{SSM2-x}). Then obviously
\begin{eqnarray}
& & K_{l}^{\bot}E_{d} = K_{l}^{\bot}K(I_{m_v}-\Phi D_{zv})^{-1}\Phi L =0 \\
& & E_{d}L_{r}^{\bot} = K(I_{m_v}-\Phi D_{zv})^{-1}\Phi LL_{r}^{\bot} =0
\end{eqnarray}
Moreover,
\begin{equation}
(K^{T}K)^{-1}K^{T}E_{d}L^{T}(LL^{T})^{-1} = (I_{m_v}-\Phi D_{zv})^{-1}\Phi
\end{equation}
which can be equivalently rewritten as
\begin{equation}
(I_{m_v}+H_{m} D_{zv})\Phi = H_{m}
\end{equation}
Therefore
\begin{equation}
(I_{m_v}+H_{m} D_{zv})^{\bot}_{l}H_{m} = (I_{m_v}+H_{m} D_{zv})^{\bot}_{l}(I_{m_v}+H_{m} D_{zv})\Phi = 0
\end{equation}
Hence, the necessity of the conditions.

On the contrary, assume that the conditions of Eq.(\ref{eqn:c4-1}) are satisfied. Then from the condition $(I_{m_v}+H_{m} D_{zv})^{\bot}_{l}H_{m} =0$, it can be declared that there exists at least one matrix, denote it by $\Phi_{0}$, such that
\begin{equation}
(I_{m_v}+H_{m} D_{zv})\Phi_{0} = H_{m}
\label{a.33}
\end{equation}

Specifically, assume that the rank of the matrices $I_{m_v}+H_{m} D_{zv}$ is equal to $r_m$. Then its singular value decomposition can in general be expressed as follows,
\begin{displaymath}
I_{m_v}+H_{m} D_{zv} =
	\begin{bmatrix}
		U_m^{[1]} & U_m^{[2]}
	\end{bmatrix}
	\begin{bmatrix}
		\Lambda_m & 0 \\ 0 & 0
	\end{bmatrix}
    \begin{bmatrix}
    	V_m^{[1]} & V_m^{[2]}
    \end{bmatrix}^T
\end{displaymath}
in which both the submatrices $U_m^{[1]}$ and $V_m^{[1]}$ have $r_{m}$ columns. Using this decomposition, it can be directly proven that
\begin{displaymath}
\left(I_{m_v}+H_{m} D_{zv}\right)^{\bot}_{l} =  U_m^{[2]T}, \hspace{0.25cm}
\left\{\left(I_{m_v}+H_{m} D_{zv}\right)^{\bot}_{l}\right\}_{r}^{\bot} =  U_m^{[1]}
\end{displaymath}

Hence, the condition $(I_{m_v}+H_{m} D_{zv})^{\bot}_{l}H_{m} =0$ is equivalent to the existence of a matrix $\overline{\Phi}_{0}$, such that
$H_{m} = U_m^{[1]} \overline{\Phi}_{0}$. Define $\Phi_{0}$ as
$\Phi_{0} = V_{m}^{[1]}\Lambda_m^{-1}\overline{\Phi}_{0} + V_{m}^{[2]}\Psi $, in which $\Psi$ is an arbitrary matrix that has a compatible dimension. Then obviously, this $\Phi_{0}$ satisfies Eq.(\ref{a.33}), no matter what value the matrix $\Psi$ takes.

Moreover, on the basis of $K_{l}^{\bot}E_{d}=0$ and $E_{d}L_{r}^{\bot}=0$, as well as the definition of the matrix $H_{m}$, straightforward algebraic manipulations show that this matrix $\Phi_{0}$ satisfies Eq.(\ref{SSM2-x}), thus proves the sufficiency of the condition.

This completes the proof.   \hspace{\fill}$\Diamond$

\begin{small}

\end{small}

\end{document}